\documentclass[10pt, preprint]{emulateapj}  

\newcommand*{\beq}{\begin{equation}}
\newcommand*{\eeq}{\end{equation}}

\newcommand*{\unit}[1]{\ensuremath{\mathrm{\, #1}}}
\newcommand*{\keV}{\unit{keV}}
\newcommand*{\pkeV}{\unit{keV}^{-1}}
\newcommand*{\ergs}{\unit{ergs}}
\newcommand*{\degree}{\ensuremath{^{\circ}}}
\newcommand*{\cm}{\unit{cm}}
\newcommand*{\km}{\unit{km}}
\newcommand*{\mum}{\unit{{\mu}m}}
\newcommand*{\ps}{\unit{s}^{-1}}
\newcommand*{\pcms}{\unit{cm}^{-2}}	
\newcommand*{\pcmc}{\unit{cm}^{-3}}	
\newcommand*{\E}[1]{\times 10^{#1}}
\def \hsi {{\it RHESSI\ }}	
\def \goes {{\it GOES\ }}
\def \soho {{\it SOHO\ }}
\def \trace {{\it TRACE\ }}

\def \Ha {H$\alpha$\ }

 \usepackage{natbib}
\begin{document}

\title{Conjugate Hard X-ray Footpoints in the 2003 October 29 X10 Flare: 
 Unshearing Motions, Correlations, and Asymmetries}


\author{Wei Liu\altaffilmark{1}\altaffilmark{2}, Vah\'{e} Petrosian\altaffilmark{2}, 
Brian R. Dennis\altaffilmark{1}, and Gordon D. Holman\altaffilmark{1}}	

\altaffiltext{1}{Solar Physics Laboratory, Code 671, Heliophysics Science Division, 
 NASA Goddard Space Flight Center, Greenbelt, MD 20771; weiliu@helio.gsfc.nasa.gov}
\altaffiltext{2}{Center for Space Science and Astrophysics, Department of Physics, Stanford University, 
  Stanford, California 94305}  

\shorttitle{Asymmetric Conjugate HXR Footpoints}
\shortauthors{Liu et al.}
\journalinfo{To appear in ApJ 2009; online at astroph http://arxiv.org/abs/0805.1055}
\submitted{Received 2008 May 05; accepted 2008 November 11}

\begin{abstract}	

We present a detailed imaging and spectroscopic study of the conjugate hard X-ray (HXR) footpoints (FPs) 
observed with {\it RHESSI} in the 2003 October 29 X10 flare.
 The double FPs first move toward and then away from each other, mainly parallel and perpendicular
to the magnetic neutral line, respectively.
The transition of these two phases of FP {\it unshearing motions} coincides with the direction reversal of the motion
of the loop-top (LT) source, and with the minima of the estimated loop length and LT height.
 The FPs show temporal {\it correlations} between HXR flux, spectral index, and magnetic field strength. 
The HXR flux exponentially correlates with the magnetic field strength, which
also anti-correlates with the spectral index before the second HXR peak's maximum,
suggesting that particle acceleration sensitively depends on the magnetic field strength
and/or reconnection rate.
 {\it Asymmetries} are observed between the FPs:
on average, the eastern FP is 2.2 times brighter in HXR flux and 1.8 times weaker in magnetic field 
strength, and moves 2.8 times faster away from the neutral line than the western FP; 
the estimated coronal column density to the eastern FP from the LT source is 1.7 times smaller.
The two FPs have marginally different spectral indexes.
The eastern-to-western FP HXR flux ratio and magnetic field strength ratio are anti-correlated only 
before the second HXR peak's maximum.
Neither magnetic mirroring nor column density {\it alone} can explain the totality of these observations, 	
but their combination, together with other transport effects, might provide a full explanation.
We have also developed novel techniques to remove particle contamination from HXR counts 
and to estimate effects of pulse pileup in imaging spectroscopy, which can be applied to other
\hsi flares in similar circumstances. 

\end{abstract}

\keywords{acceleration of particles---Sun: flares---Sun: magnetic fields---Sun: X-rays}

\section{Introduction}
\label{chp1029_intro}

In the classical picture of solar flares \citep[CHSKP,][]
{CarmichaelH1964psf..conf..451C, HirayamaT1974SoPh...34..323H, SturrockP1966Natur.211..695S, KoppR1976SoPh...50...85K},
magnetic reconnection occurs high in the corona, resulting in plasma heating and 
particle acceleration. 
Some of the high-energy particles may be trapped in the corona by plasma turbulence 
\citep[e.g.,][]{PetrosianV2004ApJ...610..550P} and/or by magnetic mirroring
\citep[e.g.,][]{MelroseDWhite1979PASAu...3..369M, LeachJ1984PhDT........33L}, where
energetic electrons produce the hard X-ray (HXR) loop-top (LT) sources \citep[e.g.,][]{Masuda1994Nature,
PetrosianV2002ApJ...569..459P, SuiL2003ApJ...596L.251S, JiangY2006ApJ...638.1140J,
BattagliaM_LTFP_2006A&A...456..751B, LiuW2006PhDT........35L}.
Some particles escape the acceleration region into interplanetary space and can contribute to
solar energetic particle (SEP) events detected at 1~AU \citep[e.g.,][]{LiuS2004ApJ...613L..81L,
KruckerS_e-spec-WIND_2007ApJ...663L.109K}.  A significant portion of the accelerated
particles escape down 	
magnetic field lines and produce bremsstrahlung HXRs (by electrons)		
or gamma-ray lines (by protons and other ions),
primarily in the {\it thick-target} \citep{BrownJ1971SoPh, PetrosianV1973ApJ...186..291P} 	
transition region and chromosphere.
This results in the commonly observed HXRs at the conjugate footpoints 
\citep[FPs;][]{HoyngP1981ApJ246L.155.HXRfp, SakaoT1994PhDT.......335S, 
PetrosianV2002ApJ...569..459P, Saint-Hilaire2008SolPh.FPasym} of the flare loop. 
Consequent energy redistribution in the lower atmosphere along the arcade of such
loops produces ribbons seen in \Ha and occasionally in white-light. 
As time proceeds, reconnection occurs at higher altitudes in the inverted-Y shaped configuration, and consequently
the HXR FPs \citep{SakaoT1994PhDT.......335S, QiuJ2002ApJ...565.1335Q, LiuW2004ApJ...611L..53L} 
and \Ha ribbons \citep{SvestkaZ1976sofl.book.....S}  	
are usually seen to move away from each other, 
while the soft and hard X-ray LT sources are observed to move upward simultaneously at a comparable speed 
\citep{GallagherP2002SoPh..210..341G, SuiL2004ApJ...612..546S, LiuW2004ApJ...611L..53L}.
At the same time, an upward growing loop system can be seen in 
\Ha \citep{ZirinTanaka.shear.1973SoPh...32..173Z},	
extreme ultraviolet \citep[EUV,][]{GallagherP2002SoPh..210..341G}, and
soft X-ray \citep[SXR;][]{Pallavicini.skylab-SXR.1975SoPh...45..411P, TsunetaS1992PASJ...44L..63T}.
%
%
Details of reconnection and particle acceleration, however, remain largely unknown.
X-ray observations of the LT and FP sources, particularly of their spatial, temporal, and spectral properties, 
combined with magnetic field measurements of the flare region,
can provide critical information about how and where electrons are accelerated subsequent to
magnetic reconnection.


Apparent motions of X-ray LT and FP sources can be understood 
as sequential excitations/formations of flare loops 	
\citep[e.g.,][]{SchrijverC.HudsonH.Oct28.2006ApJ...650.1184S} when the primary reconnection site changes its location.
Source motions deviating from the above classical picture have been observed, especially in the past decade:
 (1) The altitude of the HXR LT source was discovered by the {\it Ramaty High Energy Solar Spectroscopic Imager} 
({\it RHESSI}) to decrease during the rise of the impulsive phase
prior to the expected increase \citep{SuiL2003ApJ...596L.251S, SuiL2004ApJ...612..546S, LiuS2004ApJ...613L..81L,
LiuW_2LT.2008ApJ...676..704L}. 	
Shrinkage of loops was also observed in SXR \citep{SvestkaZ1987SoPh..108..237S,
ForbesT1996ApJ...459..330F, Vrsnak.2003Nov03shrink.2006SoPh..234..273V,		
Reeves.HinodeXRT-shrink.2008ApJ...675..868R}, EUV \citep{Li&Gan.EUV-shrink.2006ApJ...644L..97L}, 
and microwave \citep{Li&Gan.radio-shrink.2005ApJ...629L.137L} during flares, and in SXR on the quiet Sun
\citep{WangJX.loop-shrink.1997ApJ...478L..41W}.
 (2) HXR FPs exhibit even more patterns of motion. \citet{SakaoT1998opaf.conf..273S} found that in 7 out of 14 flares
observed by the {\it Yohkoh} Hard X-Ray Telescope (HXT) the FPs move away from each other, while in the rest
of the flares the FP separation decreases \citep[see also][]{FletcherL2002SoPh..210..307F} or remains roughly constant.
\citet{Bogachev.HXR-motion.2005ApJ...630..561B} extended this study to 31 {\it Yohkoh} flares and found that
only 13\% (type I) are consistent with the classical flare model showing double FPs moving away from and
nearly perpendicular to the magnetic neutral line (NL), while 26\% (type II) of the flares exhibit FPs 
moving mainly along the NL in antiparallel directions, 35\% (type III) have FPs moving along the NL in the 
same direction \citep[see also][]{KruckerS2003ApJ...595L.103K, GrigisBenz2005ApJ}, 
and the remaining 26\% show complex motion patterns. Their type II events are of particular interest as
they suggest that flare loops excited or formed more recently during reconnection are less sheared%
 \footnote{\citet{HagyardM.shear.1984SoPh...91..115H} defined the degree of magnetic shear 
 as ``the difference at the photosphere between the azimuths of a potential field 
 and the observed vector field, where the potential field is the one satisfying the boundary conditions
 provided by the observed line-of-sight field". 
 In this paper we use ``shear" to mean the angle 	
 between the line connecting the two conjugate FPs (i.e., believed to be on the same magnetic loop)
 and the normal to the NL; our ``shear angle" defined in this sense represents {\it apparent} shear,
 since we do not have accurate vector magnetic field measurements.}
\citep[e.g., see a review on {\it Hinitori} results by][]{TanakaK..Hinitori-review.1987PASJ...39....1T,
 Somov.Bastille.2002ApJ...579..863S}. 
Reduction of shear during flares has been observed for decades in the form of loops seen in 
\Ha \citep{ZirinTanaka.shear.1973SoPh...32..173Z, Rust&Bar.shear-decrease.1973SoPh...33..445R} 
and EUV \citep{AsaiA2003ApJ} with an increasing angle from the NL,
or of apparent {\bf unshearing} motions of FPs seen in \Ha \citep{AsaiA2003ApJ}, 
EUV \citep{SuYNGolubShear2007ApJ...655..606S}, 
and HXRs \citep{SakaoT1994PhDT.......335S, MasudaS2001SoPh, SchmahlE.unshear.2006ApJ...643.1271S}.
 (3) Recently \cite{JiH.converge.2006ApJ...636L.173J, JiH.unshear2007ApJ...660..893J} 
found that the decrease of the (\Ha and HXR) FP separation and shear 
occurred when the estimated height of the LT source decreases during the rise of the impulsive peak.
This has spurred renewed interest by establishing the connection between the LT descending motion 
and the FP approaching motion with decreasing shear.
LT descents reported in the past were usually observed in flares occurring near the solar limb where 
the LT height can be readily measured, but the FP motions in the east-west direction are obscured 
by projection effects. Flares close to disk center, like the one reported here,
give an alternative perspective.

{\bf Correlations} between a pair of conjugate HXR FPs are expected, since they are believed to
be produced by high-energy electrons released from the same acceleration region.
The relative timing of conjugate FPs was found to be simultaneous within an uncertainty of 0.1--0.3~s 
\citep{SakaoT1994PhDT.......335S} based on {\it Yohkoh} Hard X-Ray Telescope (HXT) observations.
For double FPs in tens of flares observed by the {\it Ramaty High Energy Solar Spectroscopic Imager} ({\it RHESSI}), 
temporal correlations in the HXR fluxes in two wide energy bands
(25--50 and 50--100~keV) with a time resolution of 8~s were identified by \citet{JinM.DingMD2007A&A}. 
Spectral correlations at individual HXR peaks were investigated by \citet{Saint-Hilaire2008SolPh.FPasym}, 
who found power-law indexes that differed by $<$0.6. 
This spectral index difference is similar to that found by 
\citet{SakaoT1994PhDT.......335S}, but smaller than the values as high as 1 or 2
reported by \citet{PetrosianV2002ApJ...569..459P}, both based on analysis of {\it Yohkoh} HXT images.

{\bf Asymmetric} FPs, i.e., conjugate FPs with different properties (HXR fluxes, magnetic field strengths, etc.)
are commonly observed \citep[e.g.,][]{SakaoT1994PhDT.......335S}.
This has been ascribed to asymmetric magnetic mirroring where a brighter HXR FP
is associated with a weaker magnetic field 	
\citep{LiJ1997ApJ, Aschwanden.FPasym.1999ApJ...517..977A, QiuJ2001ApJ, LiJP.DingMD2004ApJ606.583}. 
This picture is consistent with observations at radio wavelengths		
where brighter microwave emission appears at the FP with the stronger magnetic field 
\citep[e.g.,][]{KunduM.radioFPasym1995ApJ...454..522K, WangH1995ApJ...453..505W}.
   Exceptions to the mirroring scenario were reported by
\citet{GoffC2004A&A}, who found one third of 32 {\it Yohkoh} flares with an opposite trend, 
that is, the association of the brighter HXR FP with the stronger magnetic field.
\citet{FalewiczR2007A&A} re-examined three exceptions in the sample of Goff et al. and 
attributed this opposite asymmetry to different column densities in the two legs of the flare loop, 
as also suggested by \citet{Emslie2003ApJ0723} and \citet{LiuW2006PhDT........35L}.
Temporal variations of the flux asymmetry were found in a {\it Yohkoh} flare \citep{SiarkowskiM2004A&A},
and energy- and time-dependent variations were seen in a \hsi flare \citep{AlexanderD2002SoPh..210..323A}. 
The latter were interpreted by \citet{McClementsK2005ApJ} as a consequence of an asymmetric, energy-
and time-dependent injection of accelerated electrons.	

Previous studies of conjugate HXR FPs, in general, suffered from limited time, spatial, and/or energy 
resolution and/or coverage of HXR emission, mainly restricted by the instrumental capabilities,
 or from lack of magnetic field measurements.
We report here on a comprehensive study of the conjugate FPs in the 2003 October 29 X10 flare
observed by \hsi	 
that overcomes many of the previous shortcomings.
This flare provides a unique opportunity to track the spatial and spectral evolution of the double HXR FPs 
and their associated magnetic fields in great detail, and to study all three interrelated aspects: 
{\it unshearing motions}, {\it correlations}, and {\it asymmetries}.	
This flare occurred near disk center, where FP motions and line-of-sight magnetic field measurements have 
minimum projection effects. Its long ($\sim$20 minutes) impulsive phase and high \hsi count rates 
up to several hundred keV allow for a detailed study of variations both in time and energy. 
The flare was also well observed by the {\it Transition Region and Coronal Explorer} ({\it TRACE}), 
the {\it Solar and Heliospheric Observatory} ({\it SOHO}), and other spacecraft and many ground-based observatories.
The rich database of multiwavelength observations and a wide range of 
literature covering different aspects of this event \citep[e.g.,][]{XuY2004ApJ...607L.131X, KruckerS2005AdSpR}
are particularly beneficial for this in-depth study.

We present the observations and data analysis in \S~\ref{chp1029_obs}. These include general \hsi
light curves, images, and imaging spectroscopy.	
We investigate in \S~\ref{chp1029_FPmotion} the two phases (fast and slow) of unshearing motions of the FPs 
and the associated LT motion. In \S~\ref{chp1029_corr} we explore various correlations of the FPs,
particularly of their HXR fluxes, spectral shapes, spatial variations, and magnetic fields.
Possible contributions to the HXR flux and spectral asymmetries are discussed in \S~\ref{chp1029_asym}, followed by
our conclusions in \S~\ref{chp1029_conclude}.
A discussion of pulse pileup effects, technical details on coaligning images made by different instruments, 
a mathematical treatment of the asymmetric column density effect, and an estimate of the coronal column densities
in the legs of the flare loop are given in Appendixes~\ref{chp1029_pileup}, \ref{chp1029_coalign},
\ref{chp1029_coldep_math}, and \ref{chp1029_dens_factor}, respectively.
Note that due to the particular circumstances of this flare, we employed new procedures 
to remove particle contamination from HXR light curves and to estimate effects of pulse pileup in imaging spectroscopy.
These techniques, as a bonus of this paper, can be applied to other \hsi flares in similar difficult circumstances 
that could not be analyzed before. 

\section{Observations and Data Analysis}
\label{chp1029_obs}


We present in this section general multiwavelength and \hsi X-ray observations to indicate the
context for our detailed discussions to follow on the conjugate FPs.
The event under study occurred in AR~10486 (W5$\degree$S18$\degree$) starting at 20:37~UT on 2003 October 29,
during the so-called Halloween storms \citep[e.g.,][]{Gopalswamy2005JGRA..11009S00G}. 	
It was a {\it Geostationary Operational Environment Satellite} 
({\it GOES}) X10 class, white-light, two-ribbon flare, which produced strong 	
gamma-ray line emission \citep{HurfordG.OctNov.GR.2006ApJ...644L..93H} and helioseismic signals 
\citep{Donea.Lindsey.2005ApJ...630.1168D}. It was associated with various other solar activity, 
including a fast ($\sim$2000 $\km \ps$) halo coronal mass ejections (CMEs), and heliospheric consequences. 
This was the first white-light flare observed at the opacity minimum at $1.6 \mum$, which corresponds to the 
deepest layer of the photosphere that can be seen \citep{XuY2004ApJ...607L.131X, XuY2006ApJ}. 
There were strong photospheric shearing flows present near the magnetic NLs in this active region prior 
to the flare onset \citep{YangG2004ApJ}, which may be related to the unusually large amount of free 
magnetic energy ($\sim$6 $\E{33} \ergs$; \citealt{MetcalfT2005ApJ...623L..53M})		
stored in this AR.
 By analyzing Huairou and Mees vector magnetograms, \citet{LiuYu.Oct29Helic.2007SoPh..240..253L}
proposed that this flare resulted from reconnection between magnetic flux tubes having opposite current helicities. 
 This may be connected to the soft X-ray sigmoid structure and unshearing motions of HXR FPs
found by \citet{JiH2008ApJL.sigmoid} during the early phase of the flare. 
 \citet{LiuY2006ApJ}, using potential field extrapolations from the \soho Michelson Doppler Imager (MDI) observations,
investigated the large-scale coronal magnetic field of AR 10486 and its high productivity of CMEs.		
\citet{LiuC2006ApJ} found remote brightenings more than $2\E5 \km$ away from the main flare site.
  Solar energetic particles (SEPs) were detected after this flare by 
\goes and the {\it Advanced Composition Explorer} ({\it ACE}). 

Our goal in this paper is to understand the temporal and spectral variations of the asymmetric HXR FPs 
and their associated magnetic fields. We thus focus on HXR observations obtained by \hsi and 
line-of-sight (LOS) photospheric magnetograms obtained by \soho MDI.  
Vector magnetograms measured with chromospheric emission lines are more desirable for this
study, as relevant magnetic mirroring may take place above the chromosphere where thick-target
HXRs are produced.  
The only available vector magnetograms recorded during/near the flare time are Mees data, which, however,
has not yet been fully calibrated (K. D. Leka 2008, private communication) 	
and thus is not used here.
On the other hand, since this flare is close to disk center (W5$\degree$S18$\degree$), 
LOS MDI magnetograms provide a reasonable approximation
of the photospheric field that is assumed to be proportional
to the chromospheric field. Yet bear in mind that the field is likely tilted and can obscure
the MDI measurement due to projection effects.
It would have been interesting to examine microwave images which may show opposite FP asymmetry as
in HXRs \citep{KunduM.radioFPasym1995ApJ...454..522K}.  However, spectrograms of this flare obtained 
at the Owens Valley Solar Array \citep[OVSA][]{GaryHurford.OVSA.1990ApJ...361..290G}
do not allow for image reconstruction due to poor data quality (J. Lee \& C. Liu 2007, private communication),
while Nobeyama was not observing (before local time 6~AM).	


\subsection{Light Curves and X-ray Images}
\label{chp1029_gen_hsi}

\hsi had very good coverage of this event. However, HXR counts, particularly
at high energies ($\gtrsim$40~keV), were contaminated by counts produced by
radiation belt particles bombarding the spacecraft from all directions.
Particle-caused counts mainly influence the {\it rear} segments of \hsi detectors (unshielded), 
while the {\it front} segments are less affected during the flare since they have
a smaller volume ($\sim$0.14 that of the rear segments) and they stop most
of the downward-incident flare photons $<$150~keV \citep{SmithD2002SoPh..210...33S} that may exceed 
the number of particle counts. In addition, particle count rates
are not modulated by the grids and appear as a DC offset in the modulation pattern, which
is removed during image reconstruction \citep{HurfordG2002SoPh..210...61H}. 	
Therefore, particles do not affect our analysis in this paper since it mainly relies on images
made using front rate only. 

\begin{figure}[thbp]	
 \epsscale{1.0} 
 \plotone{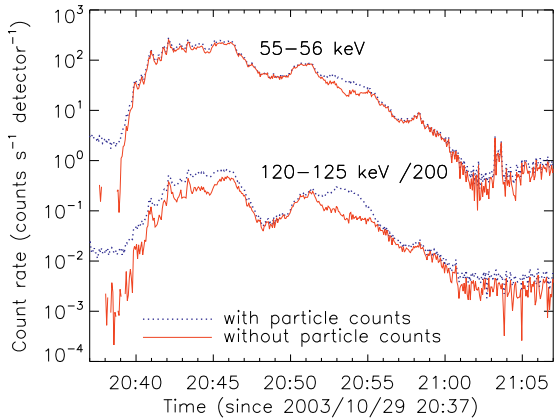}
 \caption[]
 {Count rates at two selected energies averaged over front segments 1, 3--6, 8, and 9 before ({\it dotted})
 and after ({\it solid}) the correction for particle contamination. The 120--125~keV curves are 
 scaled by a factor of 1/200.} \label{lc_no-particle.eps}
 \end{figure}

We obtained the spatially integrated X-ray fluxes free from particle contamination in the following way.	
(1) At low energies ($<$30~keV), no correction is needed since ample flare counts dominate over particle counts.
(2) At high energies (30--150~keV), we used the rear segment counts to estimate 
the contribution of particles to the front segments, since rear
counts in this energy range are almost entirely produced by particles.
To do this, we first selected a non-flare interval 20:42:40--20:47:40~UT on 2003 October 30, one day
after the flare when the spacecraft was at approximately the same geomagnetic location, to get a background estimate. 
We obtained the background count rate spectra averaged over detectors 1, 3--6, 8, and 9 for front and rear segments 
separately, and calculated the front-to-rear count-rate ratio. 	
(The ratio is close to the volume ratio $\sim$0.14 of the two segments,
because particle bombardments	
are expected to be isotropic.	
Meanwhile, the weak energy dependence of the ratio, ranging from 0.08 at 30 keV to 0.14 at 150 keV, 
may be related to the geometry of the segments.)
We then repeated this for the flare to accumulate front and rear segment count rate spectra
for every 4~s interval from 20:37 to 21:07~UT. For each time and energy bin, 
the rear count rate was multiplied by the front-to-rear ratio at this energy obtained above
and then subtracted from the front count rate. 	
A sample of the count rates before and after this correction is shown in Figure~\ref{lc_no-particle.eps}.
As expected, during the bulk of the flare duration between 20:40 and 21:00~UT, 
the estimated fractional particle contamination is minimal ($<$15\% of total counts, 
except up to 50\% between 20:52 and 20:55~UT) at a lower energy (55--56~keV),
and becomes more appreciable (up to 75\% at $\sim$20:53:30~UT) at a higher energy (120--125~keV).
Note that this technique cannot be used for energies $\lesssim$30~keV due to the threshold (lower-level discriminator)
of the rear segments set at $\sim$20~keV \citep{SmithD2002SoPh..210...33S}.
%
%

\begin{figure}[thbp]      
 \epsscale{1.1}  
 \plotone{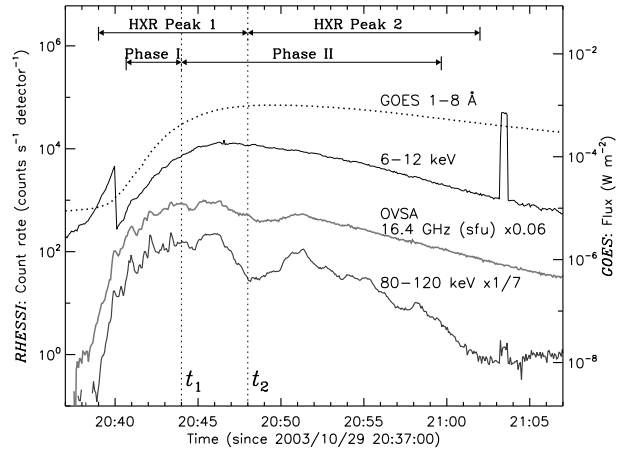}	
 \caption[Flare (X10) of 10/29/2003: {\it GOES}-12 light curves]
 {  
 X-ray fluxes from \hsi (6--12 and 80--120~keV, {\it left scale}) and {\it GOES}-12 (1--8 {\AA}, {\it right scale}), 
 and microwave flux (16.4 GHz, in sfu, {\it left scale}) from OVSA.	
 The \hsi fluxes are average count rates of front segments 1, 3--6, 8, and 9,
 with contamination from radiation belt particles removed. The artificial steps are due
 to attenuator state changes.
 The OVSA 16.4 GHz and \hsi 80--120~keV curves are scaled by factors of 0.06 and ${1 \over 7}$, respectively.
 The vertical dotted lines mark the two transitions, $t_1$ and $t_2$,	
 where $t_1$ divides Phases I and II based on FP motions and	
 $t_2$ divides two major HXR peaks (see Table~\ref{table_phases}). 	
 Count rates in the 20--50 keV range are affected pulse pileup and thus are not shown here.
 } \label{chp1029_lc.eps}
 \end{figure}

The resulting X-ray count rates 	
are also integrated in wide energy bands and are shown in Figure~\ref{chp1029_lc.eps} 
together with \goes soft X-ray and OVSA microwave fluxes.
We find that the \hsi 80--120~keV and OVSA 16.4~GHz fluxes follow each other closely in time
before 20:55~UT and both exhibit two major peaks (Peaks~1 and 2) divided at 20:48~UT.
Note that the frequency at the maximum of the OVSA microwave spectrum is $\leq$11.2~GHz 
(except for three 4~s intervals between 20:41 and 20:44~UT when it reaches up to 14~GHz),
and thus 16.4 GHz is on the optically-thin side of the spectrum which can be fitted 
with a power law.	
%
%
\begin{table}[thbp]	
\small	
\caption{Phases during the course of the flare.}
\tabcolsep 0.02in	
\begin{tabular}{lll}
\tableline \tableline
           &  Phase I               & Phase II \\
           &  20:40:40--20:44       & 20:44--20:59:40 \\
\tableline
FP unshearing motion          & fast    & slow\\
FP motion w.r.t. NL & parallel & perpendicular\\
LT altitude (estimated)       & decrease & increase\\
\tableline \tableline
           &  Peak 1                & Peak 2 \\
           &  20:39--20:48          & 20:48--21:02 \\
\tableline
mag. mirroring asymmetry   &   strong  & weak\\
column densities in loops      &   small   & large\\
FP B-fields correlation        &   exists  & disappears \\

\tableline  \end{tabular}

\label{table_phases} \end{table} 

To obtain the flare morphology and its general evolution,
we focused on a time range from 20:40:40 to 20:59:40~UT beyond which the double conjugate FPs of interest
(identified below) were not clearly imaged, due to complex morphology and/or low count rates.
We first divided this time range into 57 consecutive 20~s intervals, 
except for one interval that was shortened to 12~s to avoid the decimation state change at 20:46:36~UT.
We then reconstructed images in two broad energy bands, 12--25 and 60-100~keV,  
using the CLEAN algorithm and uniform weighting among detectors~3--8 \citep{HurfordG2002SoPh..210...61H}. 
The effective FWHM angular resolution is $9.8\arcsec$.

\begin{figure*}[thbp]      
 \epsscale{0.30}	
 \plotone{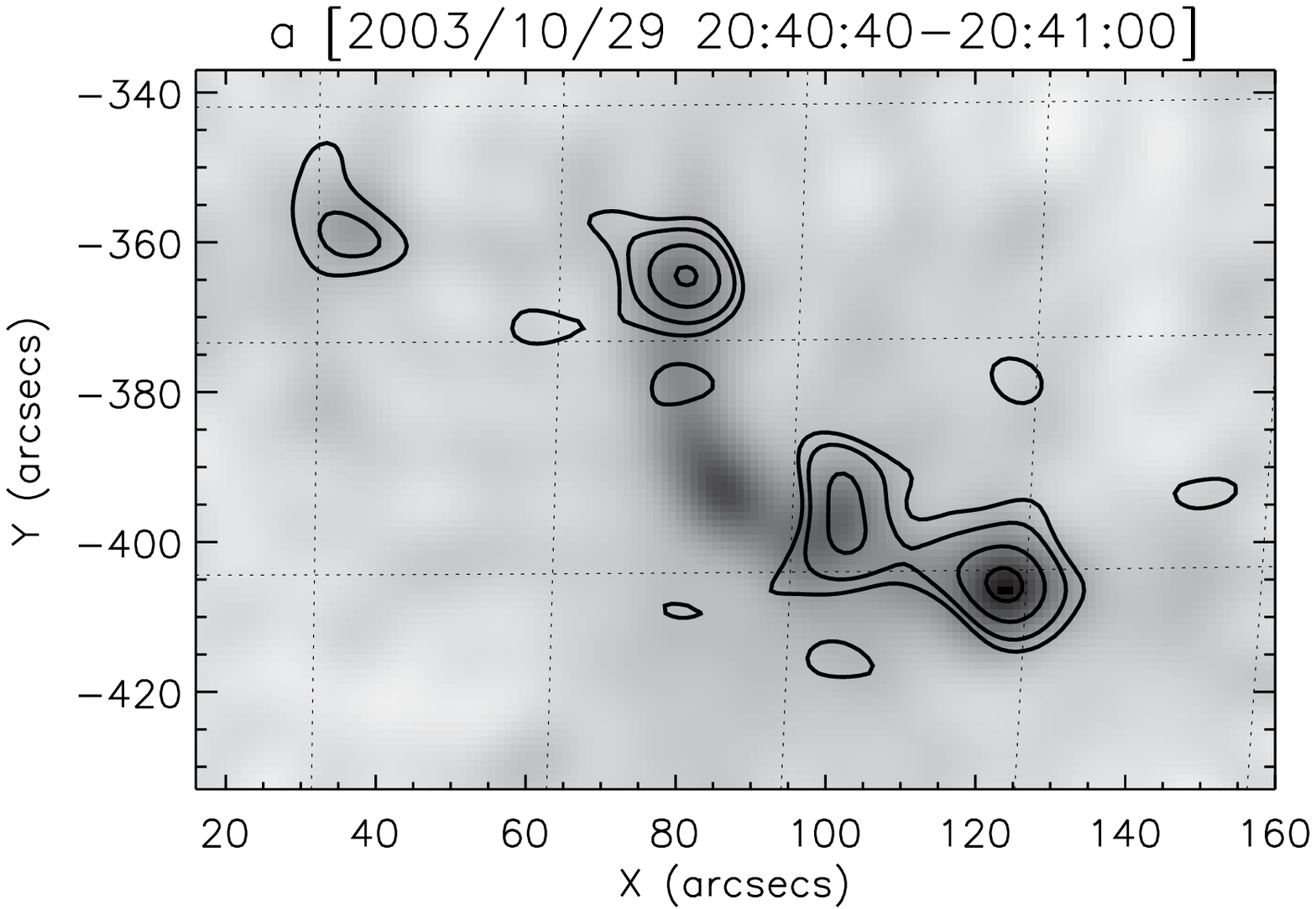}	
 \plotone{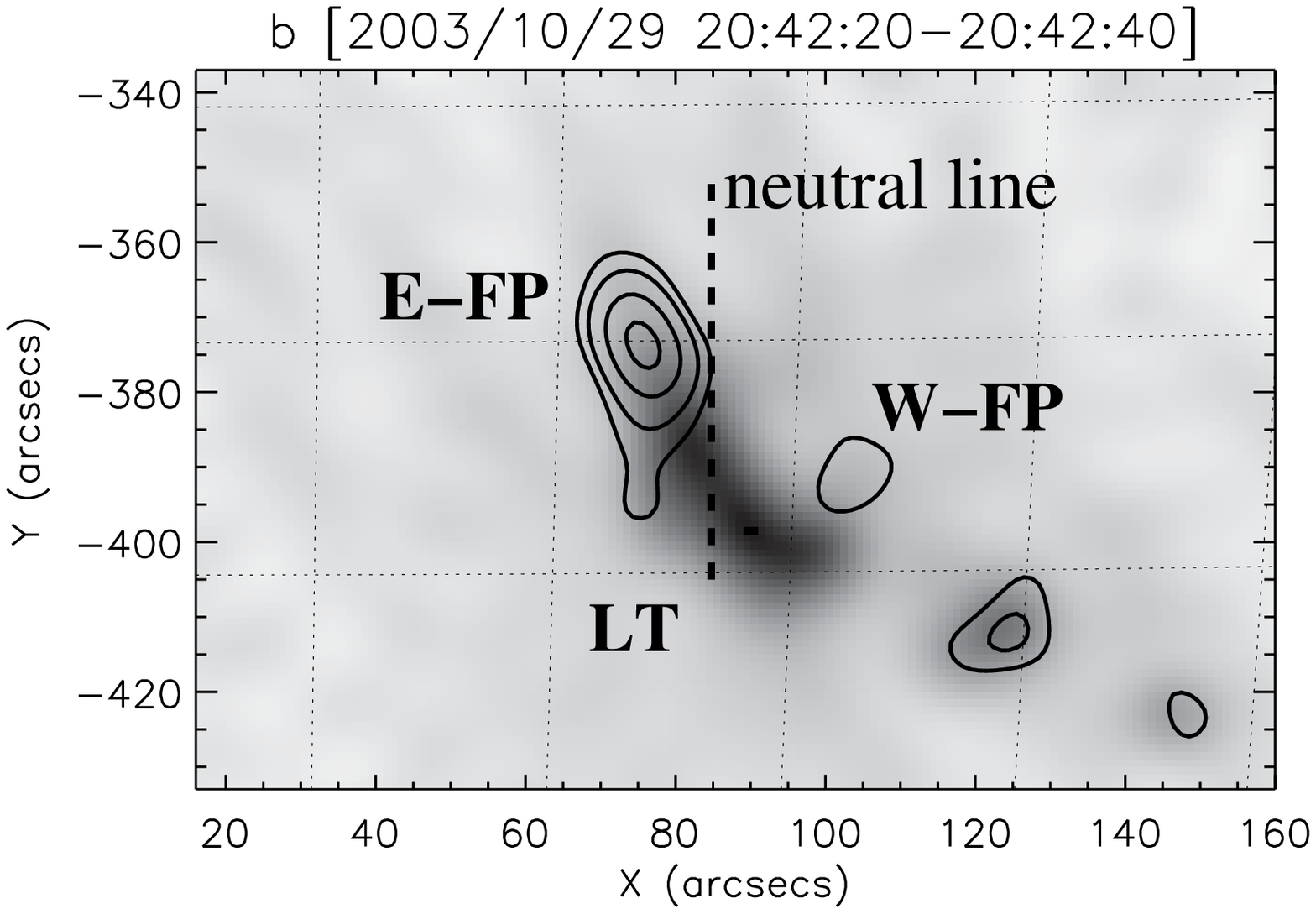}
 \plotone{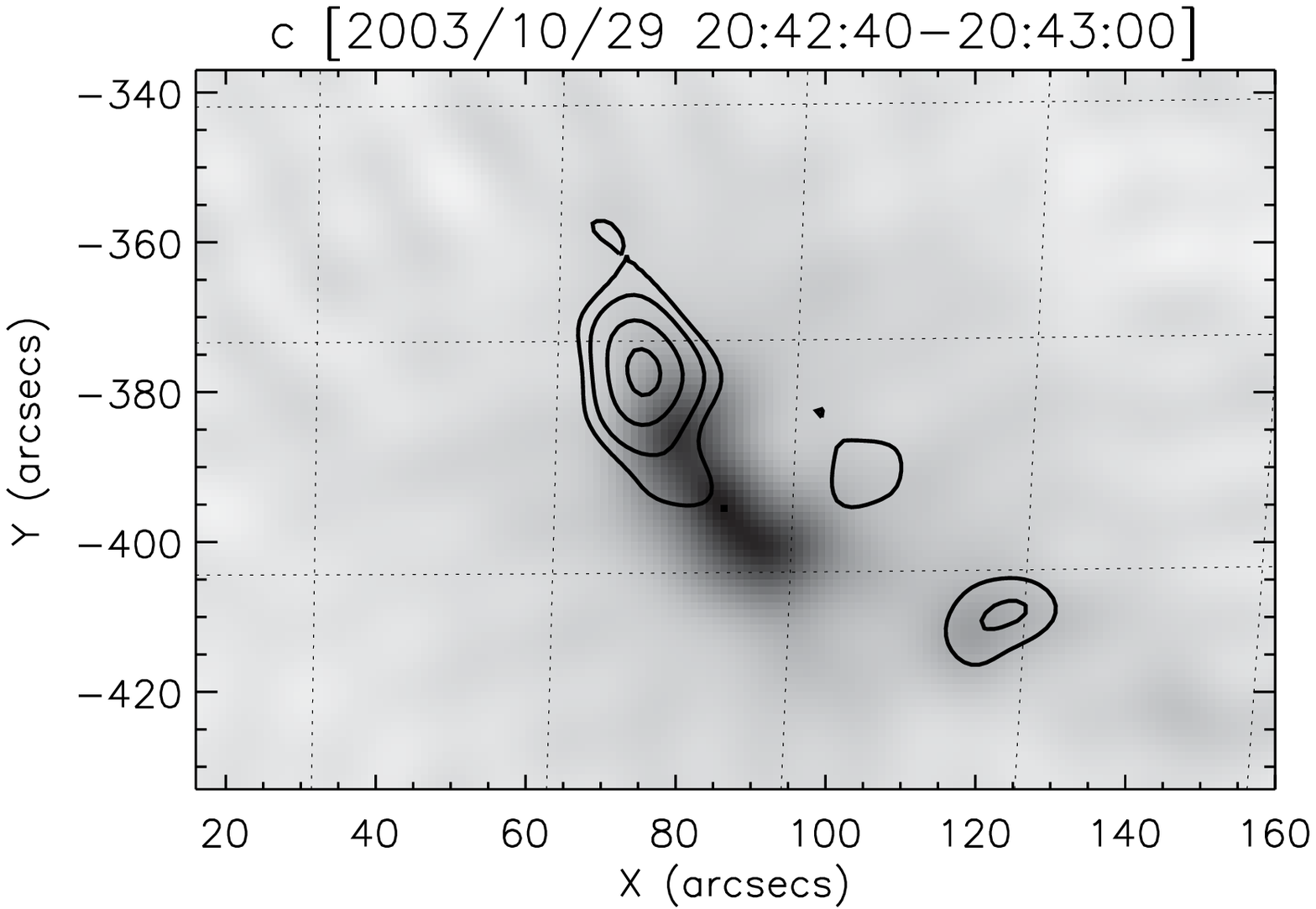}
 \plotone{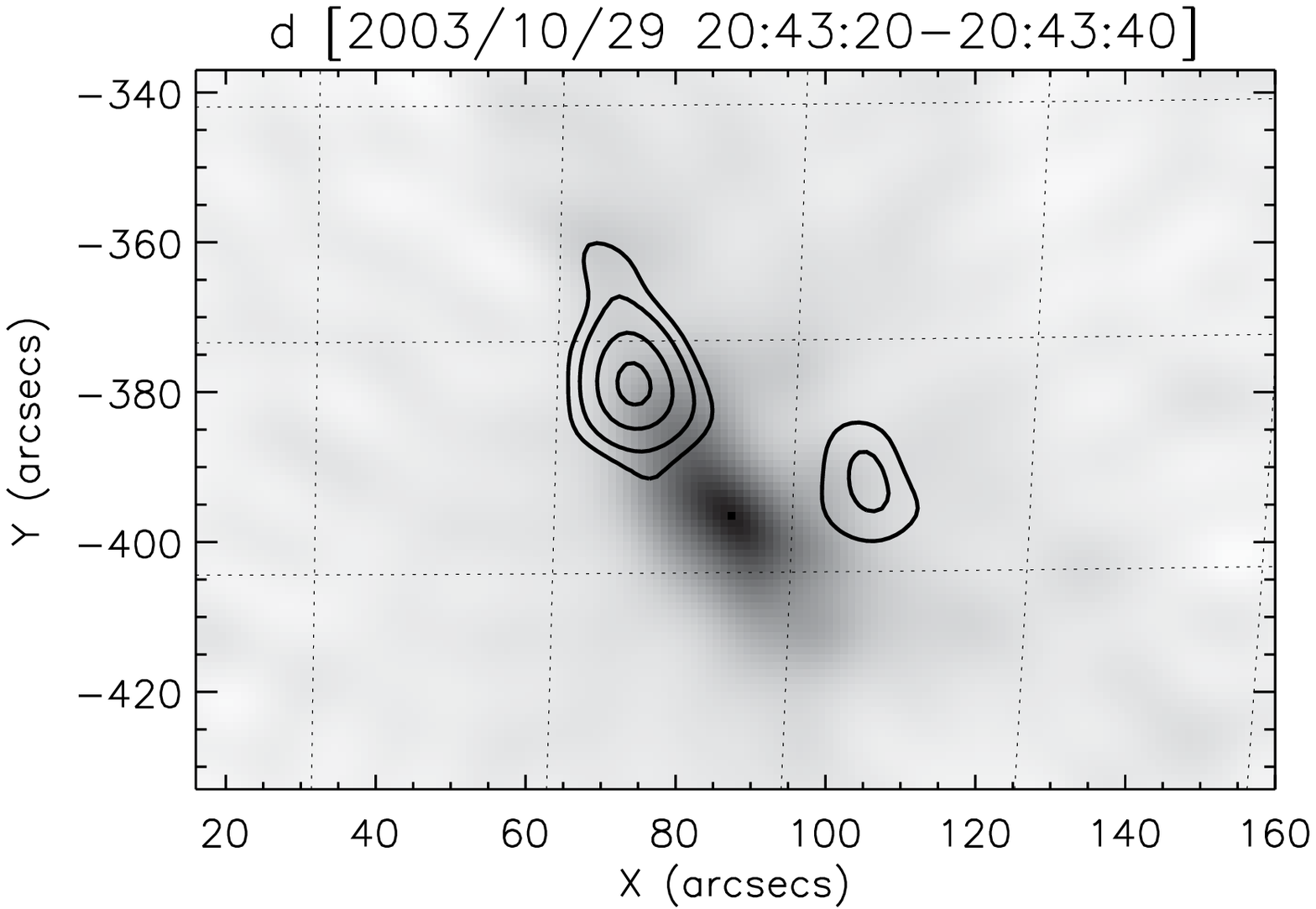}
 \plotone{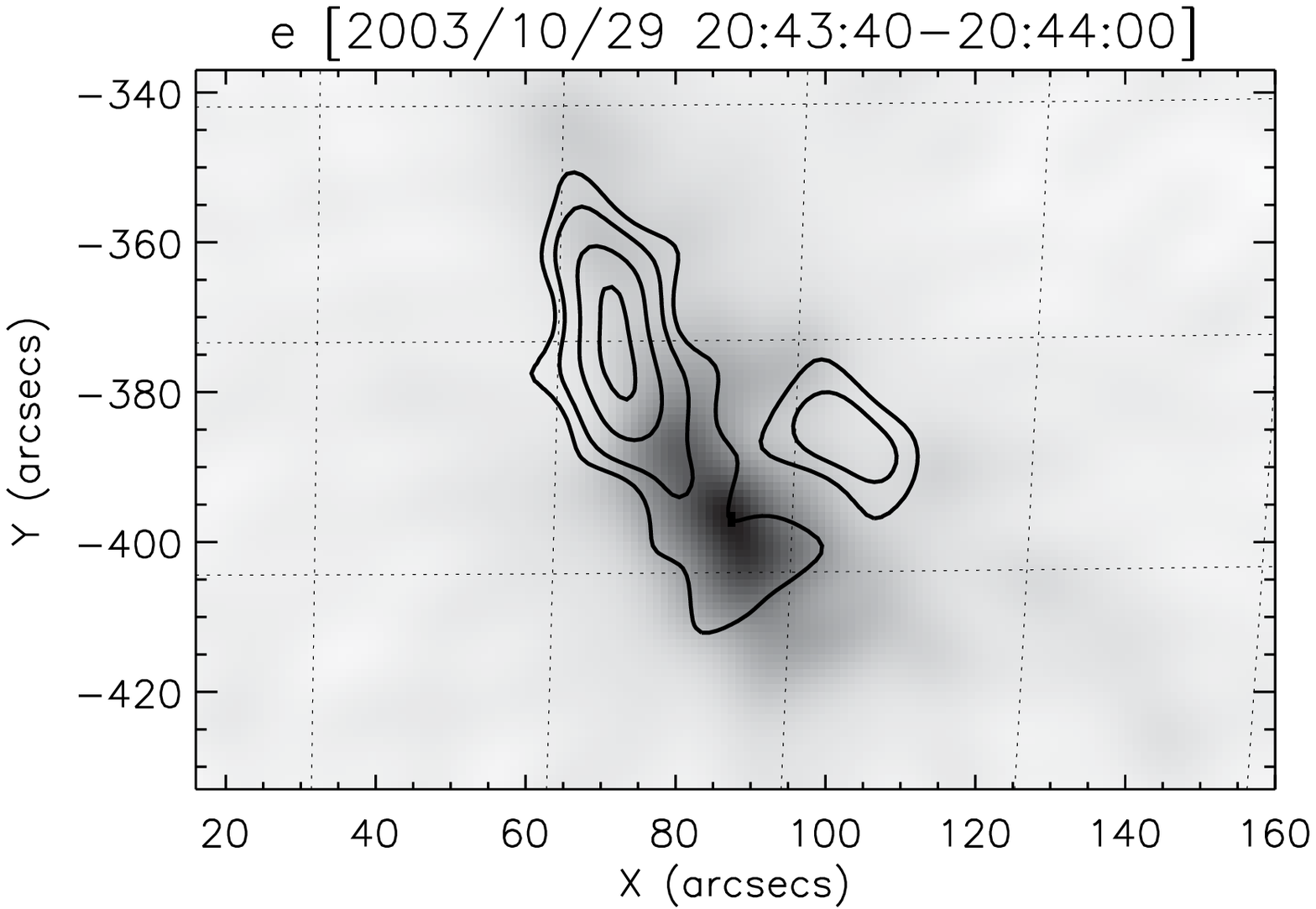}
 \plotone{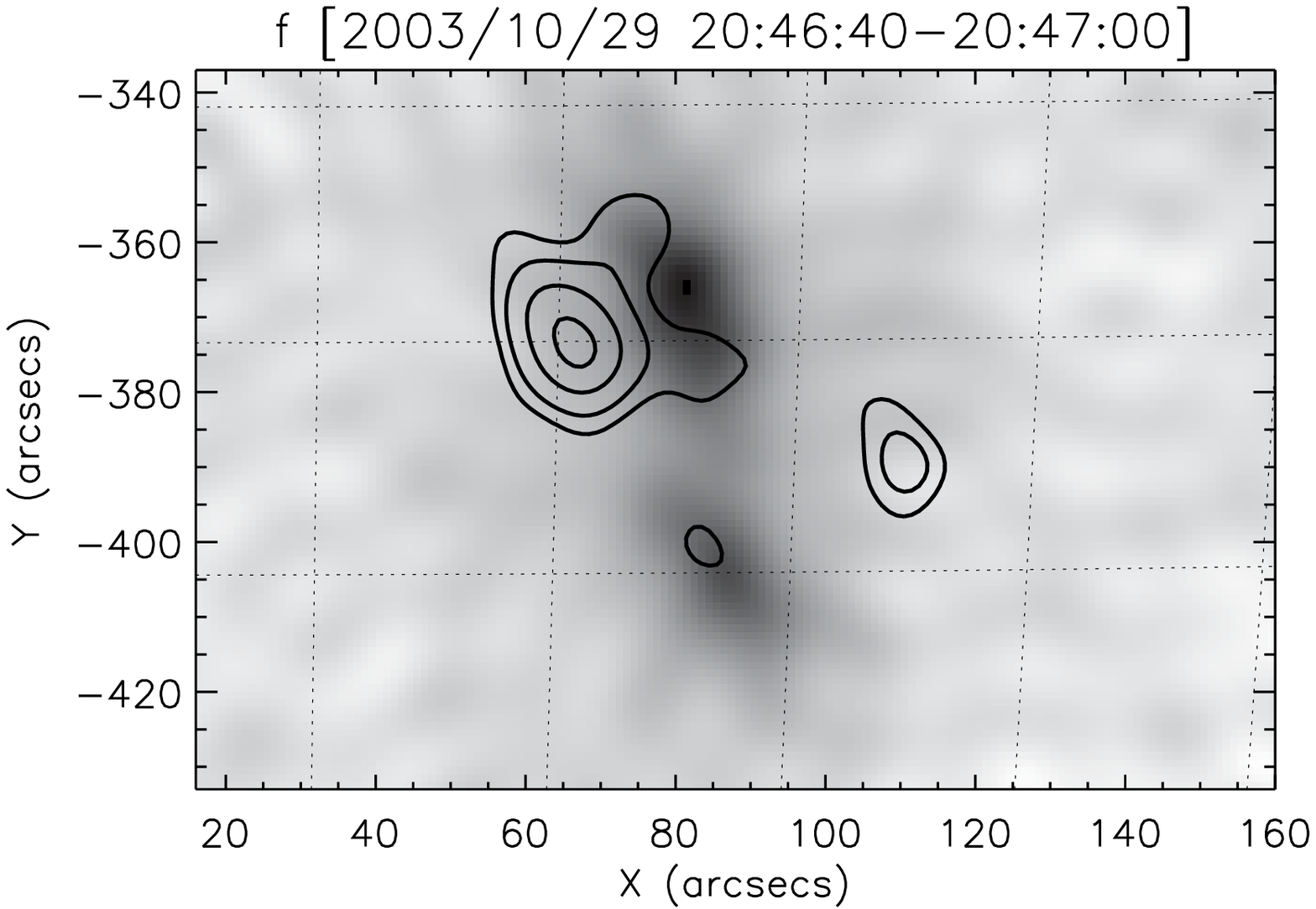}
 \plotone{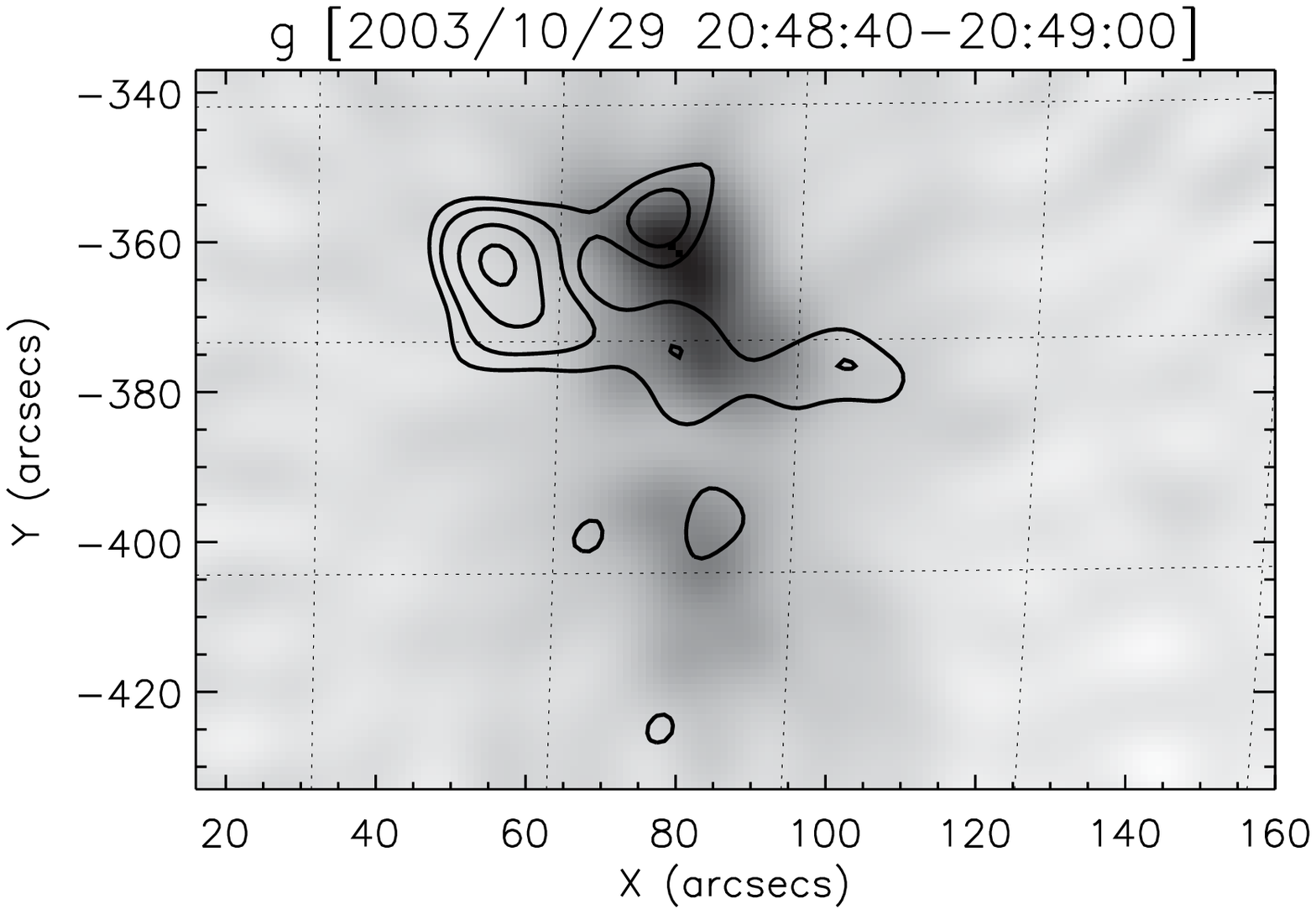}
 \plotone{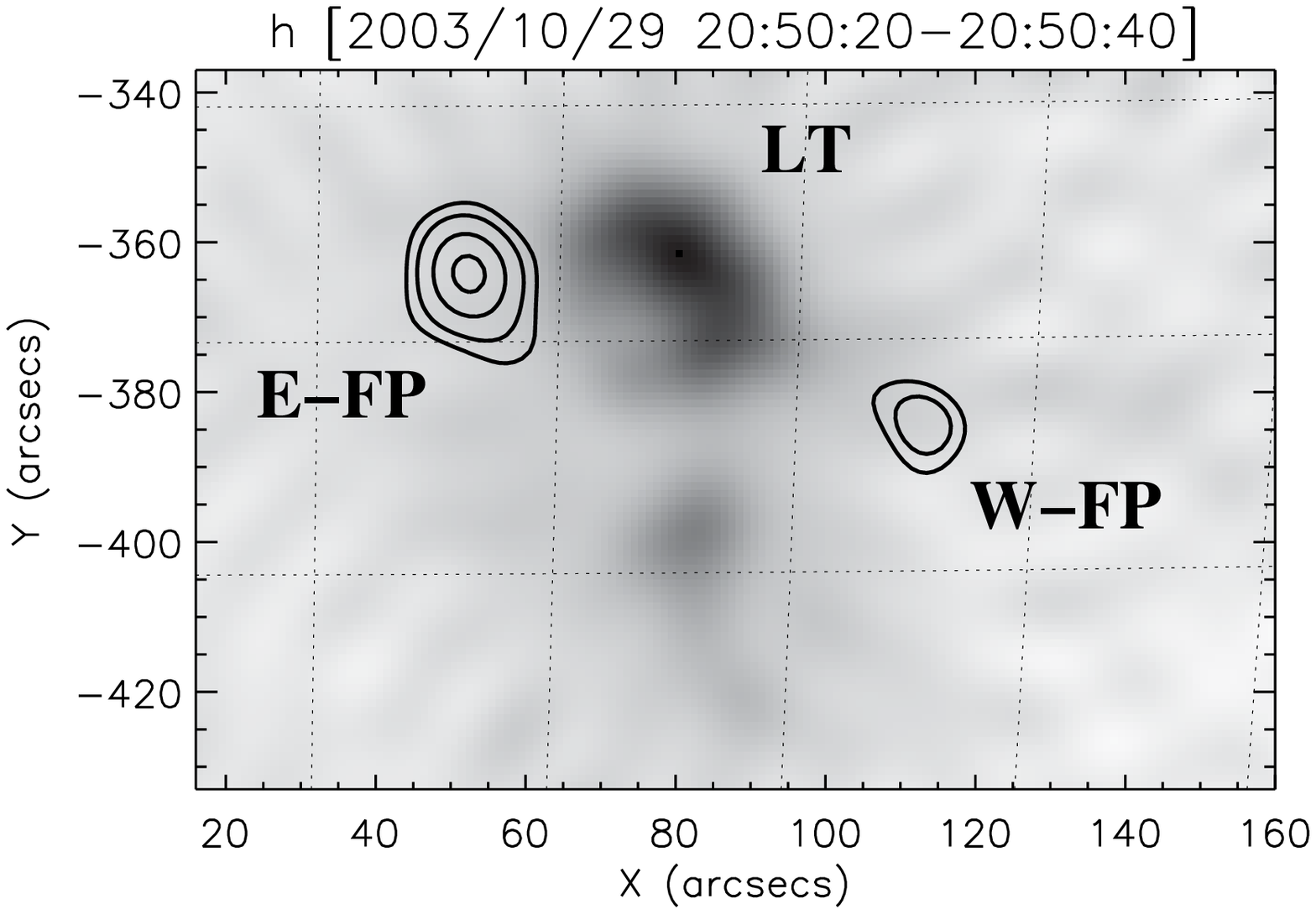}
 \plotone{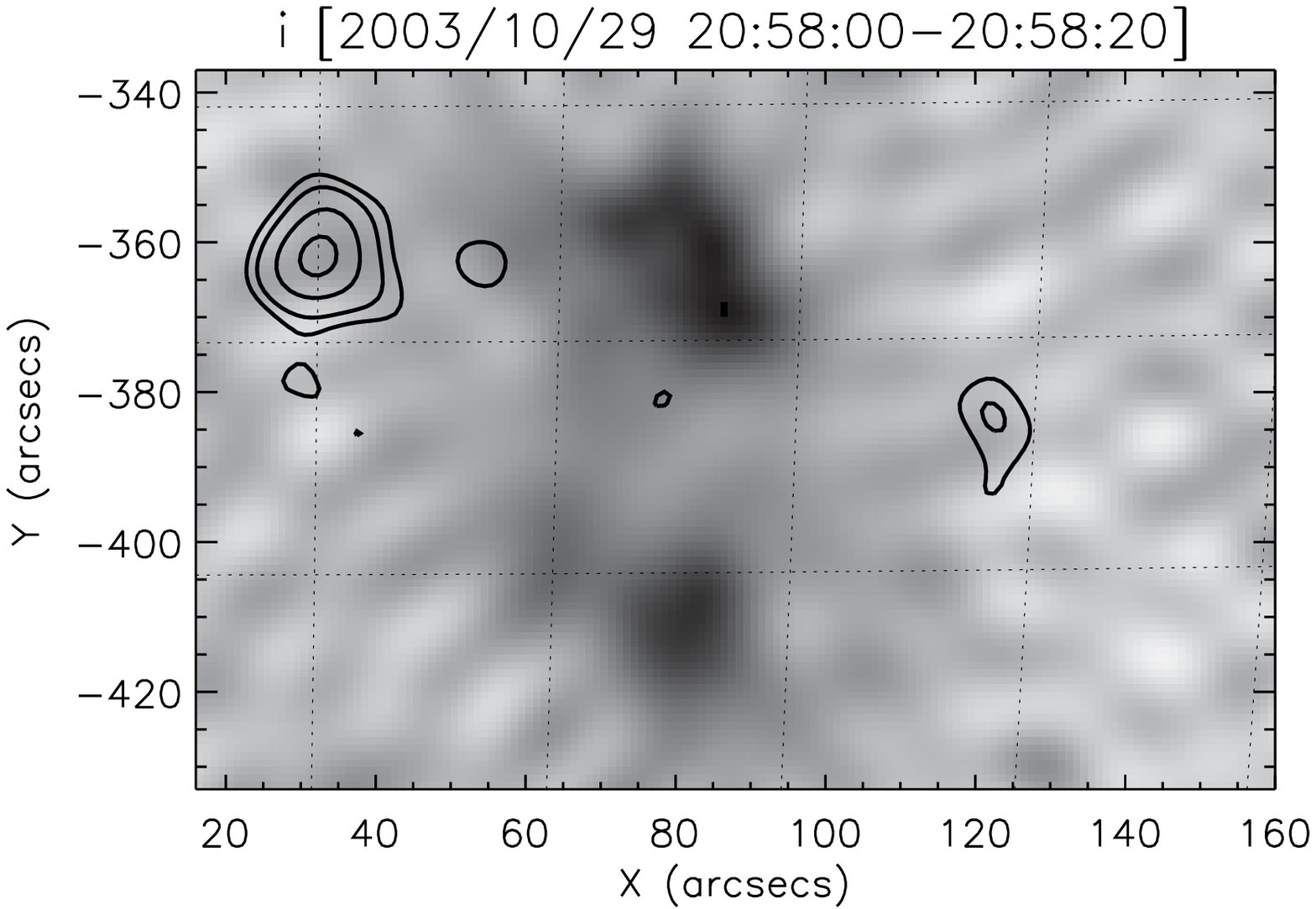}	
 \caption[\hsi 12-25 and 60-100~keV images overplotted at selected times]
 {
 Evolution of HXR sources as seen in negative \hsi CLEAN images made with detectors~3--8 (FWHM resolution $9.8\arcsec$)
 in two energy bands, 12--25~keV 
 as gray scale and 60--100~keV as contours (at 23\%, 35\%, 60\%, and 90\% of the maximum of each image).	
 The dashed line in panel {\it b} shows the	
 simplified magnetic NL also plotted in Fig.~\ref{chp1029_HSIonMDI.ps}{\it a}. 
 Correction for the solar differential rotation has been applied to this NL, and source centroids
 and images presented elsewhere in this paper when applicable unless otherwise noted.
 } \label{chp1029_hsi_img_time.eps}
 \end{figure*}

A sample of the resulting images is shown in Figure~\ref{chp1029_hsi_img_time.eps}.
Early in the flare (before 20:43:20~UT, Fig.~\ref{chp1029_hsi_img_time.eps}{\it d}), several bright points at 60--100~keV are dispersed across the image, 
suggesting FPs of multiple loops.
Part of the 12--25~keV emission appears elongated and curved 	
between the adjacent FPs, corresponding to the LT source(s). 
Toward the south-west, part of the 12--25~keV emission seems
to overlap with the FP emission, possibly due to a projection effect.	
As time proceeds, the FP structure seen at 60--100~keV becomes simpler, and only two distinct FPs are 
present (after 20:43:20~UT). They generally move away from each other. At the same time,	
the 12--25~keV emission gradually changes from one to two LT sources, one in the north and the other in the south.  
%
%

We identified the conjugate FPs and the corresponding LT source of interest as follows for detailed analysis:
(1) At later times (after 20:43:20~UT), only two FPs are seen in each image at 60-100~keV and so they are 
considered conjugate. We call the FP on the eastern (left) side E-FP and the one on the western (right) side W-FP 
(see, e.g., Fig.~\ref{chp1029_hsi_img_time.eps}{\it h}).
(2) At earlier times when more than two FPs are present, we set forth the following selection criteria:
 ({\it a}) The source morphology of the two conjugate FPs must be consistent with the picture that they are 
magnetically connected through the LT source between them seen in the corresponding 12--25~keV image
(see, e.g., Fig.~\ref{chp1029_hsi_img_time.eps}{\it b}).
 ({\it b}) During the time evolution the two FPs must show continuity and consistency in position and HXR flux,
 which other short-lived FPs lack.
Under these criteria, the selected E-FP is the brightest FP to the east of the magnetic NL 
({\it thick dashed} in Figs.~\ref{chp1029_hsi_img_time.eps}{\it b} and \ref{chp1029_HSIonMDI.ps}{\it a}), 
and W-FP is the one to the west located nearest to the NL.
(3) Once the conjugate FPs are found, their corresponding LT source was identified as the 
12--25~keV emission that lies closest to the straight line joining the FPs.
For example, at later times (see, e.g., Fig.~\ref{chp1029_hsi_img_time.eps}{\it h}),
the northern LT is selected, while the southern LT is ignored since it does not seem to have any corresponding
FP emission, presumably because of its faintness that exceeds {\it RHESSI}'s dynamic range 
\citep[$\gtrsim$10:1 for images,][p.~214 therein]{HurfordG2002SoPh..210...61H, LiuW2006PhDT........35L}.

\subsection{Imaging Spectroscopy of Footpoint and Loop-top Sources}
\label{chp1029_obs_imgspec}

\begin{figure*}[thbp]      
 \epsscale{0.8}	
 \plotone{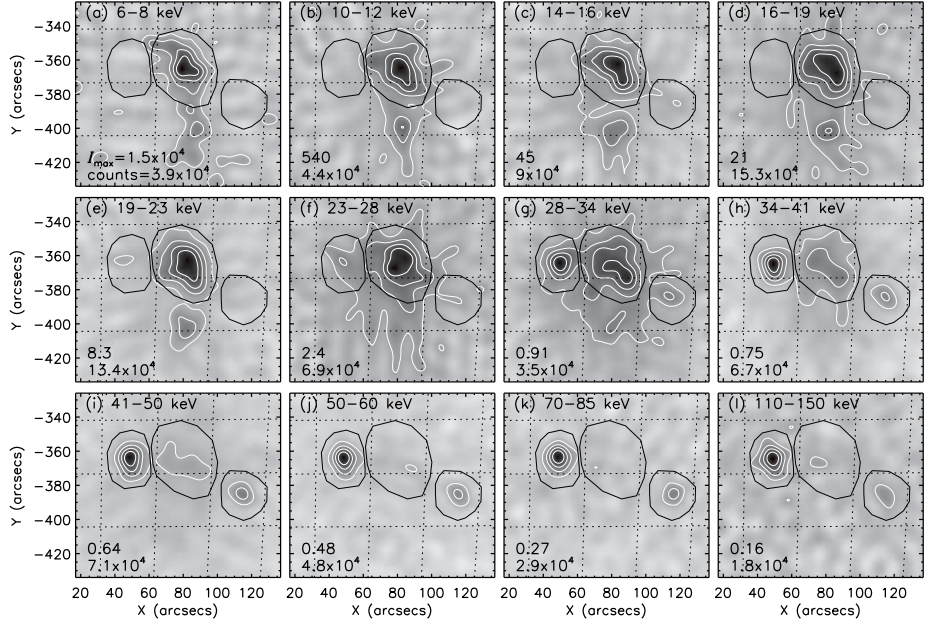}	
 \caption[]
 {
 negative CLEAN images in different energy bins at 20:51:20--20:51:40~UT made with detectors 3--9.
 The contour levels are 20\%, 40\%, 60\%, and 80\% of the maximum surface brightness, $I_\unit{max}$ 
 (shown in the lower left corner of each panel, in units of photons cm$^{-2} \ps \unit{arcsec}^{-2}$), 
 of each individual image.  The number below $I_\unit{max}$
 indicates the total counts accumulated by the detectors used.	
 The same three dark, hand-drawn polygons in each panel were used 
 to obtain the fluxes of the LT and two FP sources.
 }	\label{chp1029_img_spec_map.eps}
 \end{figure*}

Next, we examine the spectroscopic characteristics of the LT and FP sources
and their temporal evolution. For each of the 57 consecutive 20~s intervals defined above,  
we reconstructed CLEAN images in 16 energy bins that are progressively wider from 6 to 150~keV. 
A sample of these images is shown in Figure~\ref{chp1029_img_spec_map.eps} for 20:51:20--20:51:40~UT, 
where four images showing similar morphology as in neighboring energy bins are omitted.
The emission is dominated by the two LT sources at low energies and the double FP sources 
at high energies.  

The next step was to obtain photon fluxes of the sources for each time interval. 
For each FP source, we used a hand-drawn polygon
that envelops all the 10\% (of the maximum brightness of the image) contours at 	
energies where this FP source was clearly imaged. For the corresponding LT source, we drew a polygon that 
encloses the 20\% contours, 	
which was selected to minimize spatial contamination from the FPs.  
We then read the resulting multiple-energy image cube into the standard
\hsi spectral analysis software (OSPEX) package. 
This package integrates photon fluxes inside each polygon, and uses the full detector response 
matrix to estimate the true incident photon spectrum.
The RMS of the residual map of the CLEAN image was used to calculate the uncertainty for the photon flux
in each energy bin,		
with proper consideration of the source area and grid spatial resolution. 
This imaging spectroscopy technique is detailed by \citet{LiuW_2LT.2008ApJ...676..704L}.
Note that we did not use contours at a fixed level (as opposed to polygons fixed in space) 
to obtain the fluxes because of the complex source morphology that makes such contours	
vary with energy.

One important issue for this X10 flare is pulse pileup \citep{SmithD2002SoPh..210...33S} that at high count 
rates distorts the count-rate spectrum. 
We have discussed in Appendix~\ref{chp1029_pileup} various effects of pileup on our analysis and the remedy that we
have applied to minimize them.  Although it is currently not possible to obtain accurate spectra
throughout the full energy range for all sources, 	
pileup mainly affects the LT spectra in the range of 20--50~keV 
(e.g., see Figs.~\ref{chp1029_img_spec_map.eps}{\it h} and \ref{chp1029_img_spec_map.eps}{\it k}).
In other words, pileup effects on spectral shapes are negligible for 
the LT sources below 20~keV and for the FP sources above 50~keV. 
This conclusion enabled us to confine the extent of pileup effects both in
energy and in space.	
 We thus fitted the LT spectrum below 20~keV with an assumed isothermal model from
CHIANTI ver.~5.2 \citep{YoungP2003ApJS.CHIANTI}, using the default coronal iron abundance 
of 4 times the photospheric value,
to determine its temperature ($T$) and emission measure (EM);
we fitted the FP spectrum above 50~keV with an assumed single power-law model to find its spectral
index ($\gamma$) and normalization flux ($I$) at the reference energy of 50~keV.  
%
%
%
 \begin{figure}[thbp]      
 \epsscale{1}  
 \plotone{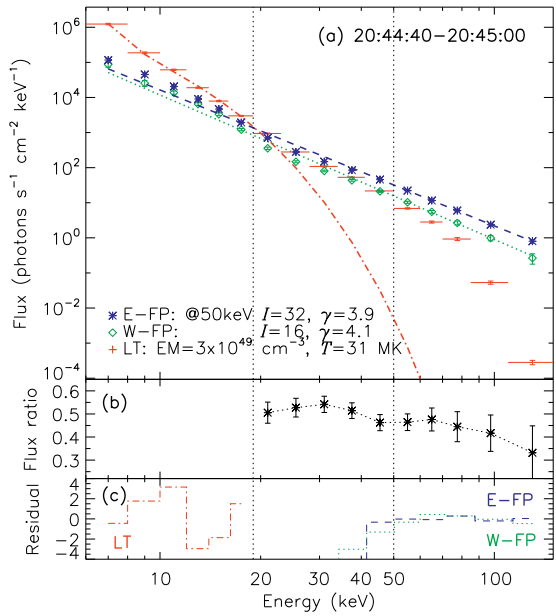}	
 \caption[]
 {
 ({\it a}) Spatially resolved spectra of the LT and two FP sources at 20:44:40--20:45:00~UT.	
 The horizontal bars on the LT spectrum represent the energy bin widths in the range of 6--150~keV.
 The dashed and dotted lines are power-law fits from 50--150~keV for E-FP and W-FP, respectively,
 and the dot-dashed line is a single temperature thermal fit from 6--19~keV for the LT.
 The legend shows the photon fluxes ($I$) at 50~keV and the spectral indexes ($\gamma$) 
 for the FPs, and the emission measure (EM) and temperature ($T$) for the LT.	
  ({\it b}) Ratio of the W-to-E FP fluxes.
  ({\it c}) Fitting residuals normalized by the 1$\sigma$ uncertainties for the LT ({\it dot-dashed}), E-FP ({\it dashed})
 and W-FP ({\it dotted}) sources.
 }	\label{chp1029_img_spec.eps}
 \end{figure}
%
%
 \begin{figure}[thbp]      
 \epsscale{1}  
 \plotone{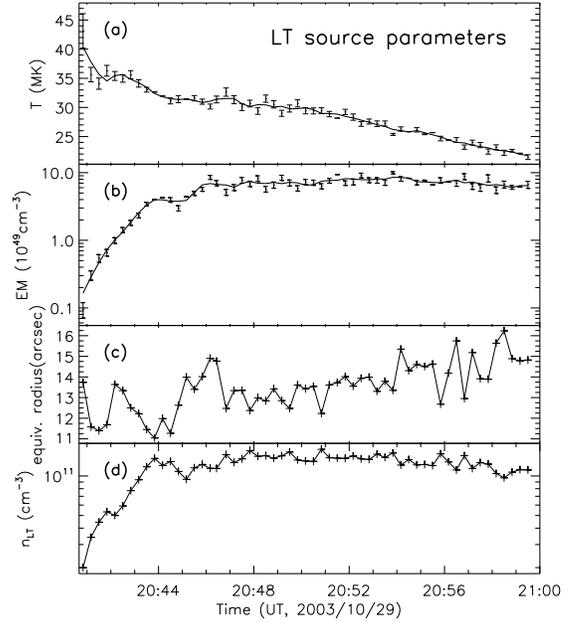}	
 \caption[]
 {
 Evolution of spectroscopic parameters of the LT source.
 ({\it a}) Temperature $T$ and ({\it b}) emission measure EM with 20~s integration time.
 The solid lines are 7-point box-car smooths of the original fitting results indicated by the symbols. 
 ({\it c}) Radius $r$ of the equivalent sphere of the LT source (see \S \ref{chp1029_obs_imgspec}) and
 ({\it d}) corresponding electron number density $n_{\rm LT}$.
 }	\label{chp1029_img_spec_time.eps}
 \end{figure}
%

Spectra of the LT and FP sources are shown in Figure~\ref{chp1029_img_spec.eps}{\it a} for the interval of 
20:44:40--20:45:00~UT (during the main impulsive peak). 
Above 50~keV, both FP spectra have a power-law shape, with the E-FP flux being twice that of W-FP but only slightly harder.
 Consequently, the W-to-E ratio of the two FP spectra generally decreases with energy
(Fig.~\ref{chp1029_img_spec.eps}{\it b})	
or stays constant within uncertainties.		
Below 20~keV the LT spectrum shows the exponential shape of isothermal bremsstrahlung emission, 
with the iron line feature at 6.7~keV visible. Note that below 50~keV
the FP spectra may be compromised by	
pileup effects%
 \footnote{We note that the same power-law trends of the two FP spectra 	
 extend below 50~keV down to $\sim$30~keV, suggesting that 	
 pulse pileup  may have minimal effects on the spectral shapes of the FPs, and that our selection 	
 of 50~keV as the lower limit for reliable FP spectra is likely to be unnecessarily conservative.
 }
and spatial contamination from the LT source, and likewise above 20~keV the apparent LT flux is 
contaminated by FP emission at the same energy and by pileup from lower energies
(Fig.~\ref{chp1029_img_spec.eps}{\it a}).	
We show in Figure~\ref{chp1029_img_spec_time.eps} the spectral evolution of the LT source 
and defer that of the FP sources to \S~\ref{chp1029_corr_spec}.

  In order to infer the density of the LT source, we assumed that it has a spherical shape
with the projected area $a$ 
equal to the area inside the 50\% brightness contour at 12--25~keV.
 We then obtained the radius $r=(a/\pi)^{1/2}$ and 
volume $V=4\pi r^3/3$ of the {\it equivalent sphere} and the corresponding density $n_{\rm LT}=[\unit{EM}/(V f)]^{1/2}$.
In doing so we assumed a filling factor $f$ of unity, 
which means the density obtained here would be a lower limit, 
and used the EM values smoothed with a 7-point box-car to minimize fluctuations possibly
caused by the inevitable anti-correlation between $T$ and EM during spectral fitting.	
The values of $r$ and $n_{\rm LT}$ as functions of time are shown in
Figures \ref{chp1029_img_spec_time.eps}{\it c}		
and \ref{chp1029_img_spec_time.eps}{\it d}, respectively.
As evident, the size of the sphere stays roughly constant between $11\arcsec$--$16\arcsec$
and thus the density follows the same trend as the EM. 
%

%

\section{Two-phase Footpoint Unshearing and Loop-top Motions}
\label{chp1029_FPmotion}

\begin{figure}[thbp]      
 \epsscale{1}	
 \plotone{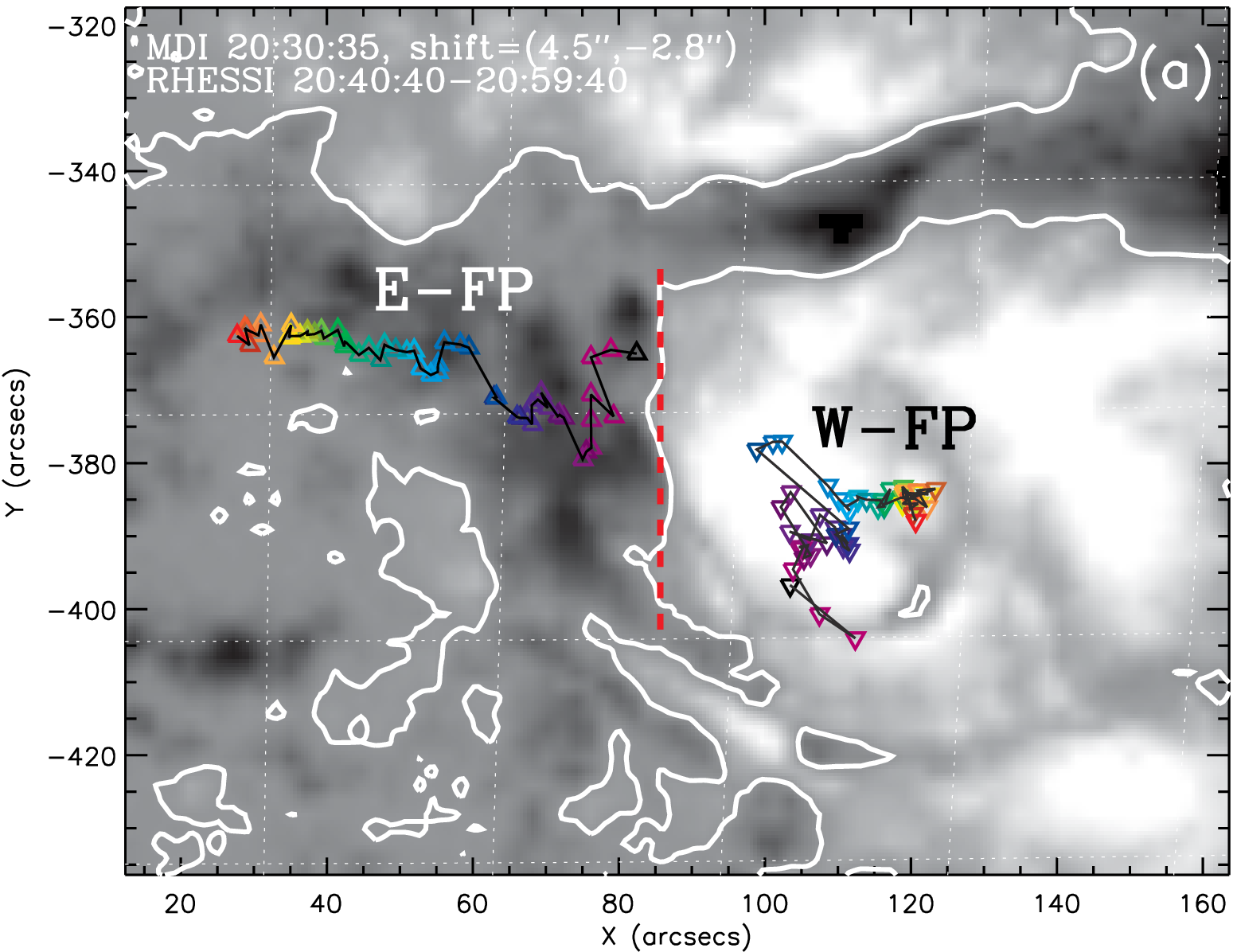}
 \plotone{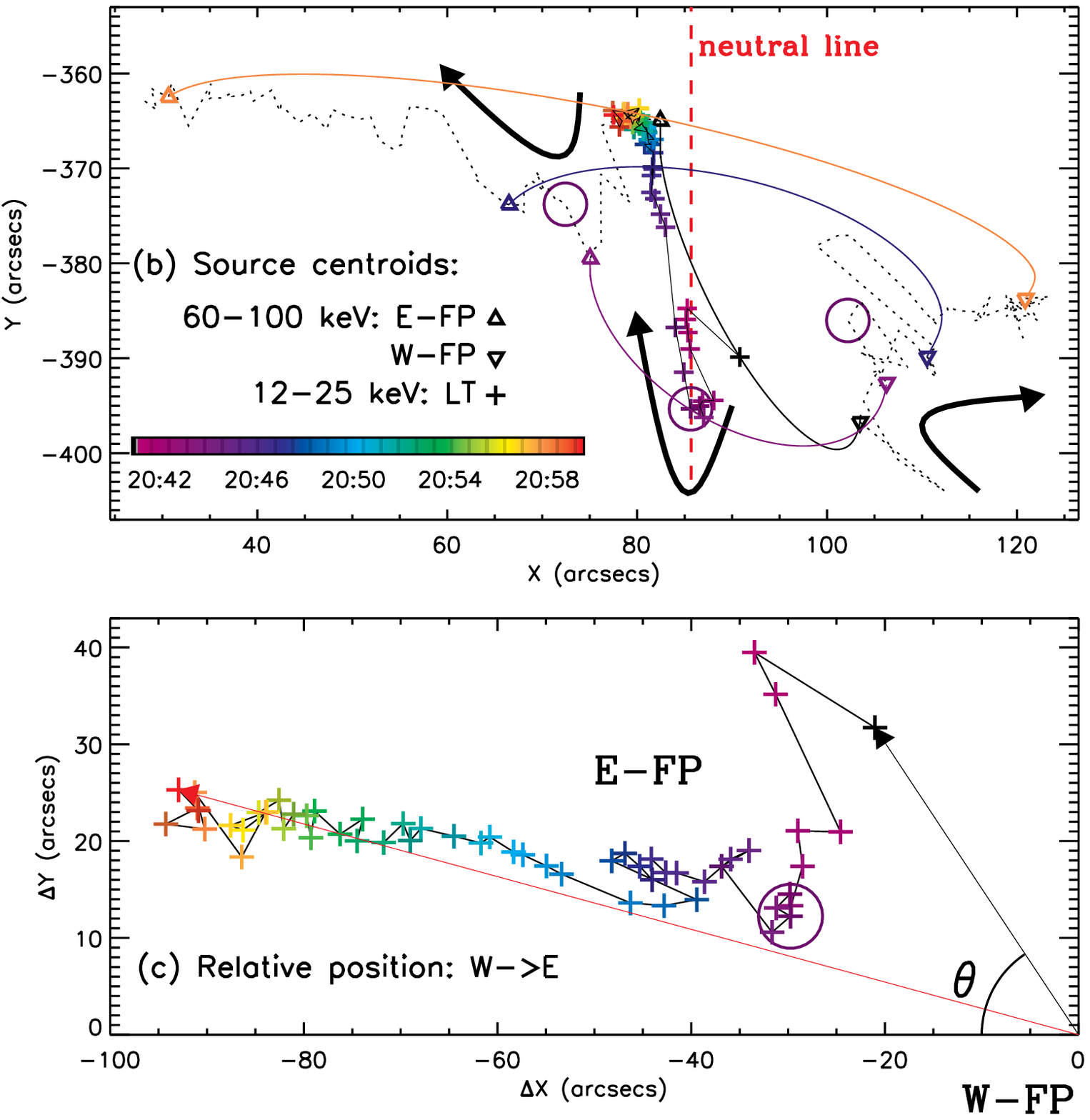}		
 \caption[MDI magnetogram and FP locations]
 {  
   ({\it a}) preflare \soho MDI magnetogram (centered at 20:30:35~UT with an integration time of 30~s) 
 overlaid with magnetic NLs in white and the centroids of \hsi FPs (E- and W-FP, at 60--100~keV) in color.  
 The white (dark) gray scale represents positive (negative) line-of-sight magnetic fields pointing away 
 (toward) the observer. The vertical red dashed line here and in ({\it b}) is the	
 simplified NL that follows the observed magnetic NL between the two FPs.
 The temporal evolution at 20~s intervals of the \hsi centroids is indicated by the color bar in ({\it b}).
   ({\it b}) Evolution of the centroid of the \hsi LT source at 12--25~keV ({\it plus signs}).
 The tracks of the FP centroids shown in ({\it a}) are repeated here as dotted lines, along with
 positions at four selected times (labeled {\it a}, {\it d}, {\it f}, and {\it i} 
 in Fig.~\ref{chp1029_hsi_img_time.eps}) marked by triangles.
 We show semi-circular model loops at these times projected onto the sky plane, 
 each of which connects three centroids (of two FPs \& one LT) of the same color.
 The thick dark arrows indicate the general direction of motion for the LT and two FPs. 
   ({\it c}) Relative centroid positions of E-FP with respect to W-FP which is selected as the origin.
 The start and end of the time evolution are marked by the black and red arrows.	
 $\theta$ is the shear angle between the normal (due west) to the NL and the line joining the 
 two FP centroids.	
 The open	
 circle here and those in ({\it b}) 
 mark the transitional time $t_1$=20:44~UT from fast to slow unshearing motions of the FPs (see \S \ref{chp1029_FPmotion}).
 } \label{chp1029_HSIonMDI.ps}
 \end{figure}

We now examine in detail the spatial evolution of the double FP sources and the corresponding LT source,
by tracking the migration of their emission centroids. For each 12--25 image obtained in \S~\ref{chp1029_gen_hsi} 
we used a contour at 50\% of the maximum brightness of the LT source to locate its 	
centroid, while for each 60--100~keV image we used a 90\% contour of each conjugate FP.
The reason for a higher contour level for the FPs (than the LT)
is that the E-FP source spreads along the flare ribbon (see, e.g., Fig.~\ref{chp1029_hsi_img_time.eps}{\it e}) 
and we need this 	
brightest ``kernel" to obtain the corresponding magnetic field strength at the FP 	
(see \S \ref{chp1029_corr_B}). The resulting centroids are shown in Figure~\ref{chp1029_HSIonMDI.ps}.


The background preflare MDI magnetogram in Figure~\ref{chp1029_HSIonMDI.ps}{\it a} was corrected 
from \soho L1 view to Earth view and shifted by $\Delta x=4.5 \arcsec \pm 2.0 \arcsec$ 
and $\Delta y=-2.8 \arcsec \pm 2.0 \arcsec$ in the solar east-west ($x$) and south-north ($y$) 
directions, respectively, to match the \hsi aspect believed to be have sub-arcsecond accuracy
\citep{FivianM2002SoPh..210...87F}. The required shifts were determined by cross-correlating
MDI magnetic anomaly features \citep[e.g.,][]{Qiu.Gary.BfieldAnomaly2003ApJ.599.615, SchrijverC.HudsonH.Oct28.2006ApJ...650.1184S} 
with HXR FPs, as described in Appendix~\ref{chp1029_coalign}. 
The MDI map and all \hsi centroids were corrected for the solar rotation 
and shifted to their corresponding positions at a common time (20:50:42~UT) in the middle of the flare.
As is evident, E-FP is located in the negative ({\it dark}) polarity to the left of the 
simplified magnetic NL ({\it red dashed}), while W-FP is in the positive ({\it white})
polarity to the right.

\begin{figure}[thbp]      
 \epsscale{1} 
\plotone{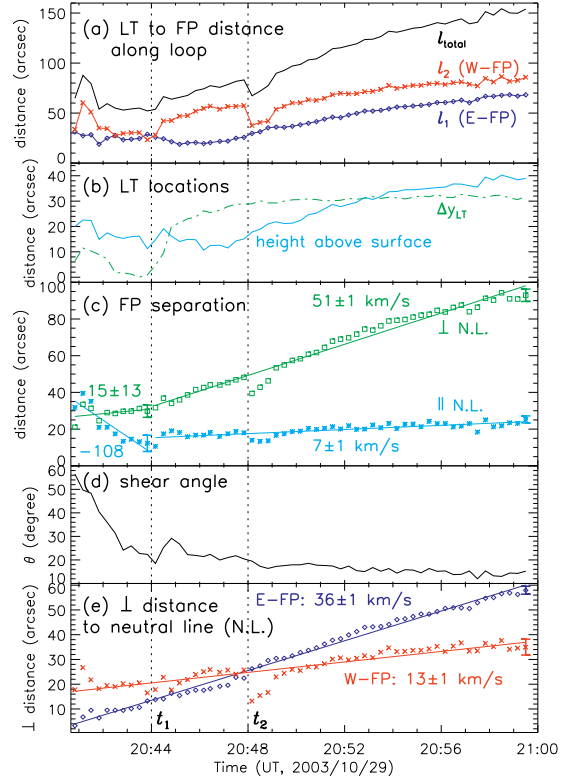}	
 \caption[]
 {
  ({\it a}) History of distance ($l_1$ and $l_2$) from the LT centroid to the centroids of the two FPs along
 the semi-circular model loop as shown in Fig.~\ref{chp1029_HSIonMDI.ps}{\it b}, together with the length of the loop 
 ($l_{\rm total}= l_1 + l_2$).
   ({\it b}) Relative displacement ($\Delta y_{\rm LT}$, {\it dot-dashed})
 of the LT centroid 
 parallel to the magnetic NL shown in Fig.~\ref{chp1029_HSIonMDI.ps}{\it b} 
 and estimated height of the LT centroid above the solar surface ({\it solid}). 
   ({\it c}) Orthogonal components of the separation between	
 W-FP and E-FP perpendicular ({\it squares}) and parallel({\it asterisks})  
 to the NL. 
 The straight lines are linear fits to the distances in the two phases, labeled with the corresponding
 velocities (in $\km\ps$).
   ({\it d}) Shear angle $\theta$ defined in Fig.~\ref{chp1029_HSIonMDI.ps}{\it c}.
   ({\it e}) Perpendicular distances to the NL from E-FP ({\it diamonds}) 
 and W-FP ({\it crosses}). Linear fits for the whole flare duration are shown as the straight lines.
 The error bar shown on the last data point of each line here and in ({\it c}) is	
 the RMS deviation of the data from the corresponding fit.	
 }	\label{chp1029_fp_motion.eps}
 \end{figure}

In an attempt to correct for projection effects and to obtain the true 3D loop geometry, we assumed that 
the centroids of the LT and two FP sources at a given time are connected by a semi-circular loop.
We then used the solar $x$ and $y$ coordinates of these three points in the sky plane
to determine the size and the orientation of the semi-circle in 3D space,
knowing that the FPs are located on the solar surface 
and the LT in the corona.  A sample of the loops at selected times	
is shown in Figure~\ref{chp1029_HSIonMDI.ps}{\it b}. 
We find that the inclination angle between the model loop and the vertical direction 
ranges from $14\degree$ to $63\degree$,              
and that the loop length ($l_{\rm total}$; see Fig.~\ref{chp1029_fp_motion.eps}{\it a}) 
generally first decreases and then increases with a minimum at 20:43:50~UT.

The LT centroids ({\it plus signs}) as shown in Figure~\ref{chp1029_HSIonMDI.ps}{\it b}
are situated at all times close to 	
the NL ({\it red dashed}) as expected,
and form two clusters, one in the south and the other in the north.
As time proceeds, the LT centroid appears to move from the apex of one loop to another along the arcade 
seen in \trace 195~\AA\ (not shown).	
It first gradually moves southward until 
20:43:30~UT (marked by the middle circle in Fig.~\ref{chp1029_HSIonMDI.ps}{\it b}), 
when it starts to rapidly shift to the northern cluster and then continue moving northward at progressively lower speeds.
This can be more clearly seen from its relative displacement projected onto	 
the north-south NL as a function of time (Fig.~\ref{chp1029_fp_motion.eps}{\it b},
$\Delta y_{LT}$).	
The height of the LT centroid estimated from the model loops is also shown in Figure~\ref{chp1029_fp_motion.eps}{\it b}
and exhibits a general decrease followed by an increase.%
%
%

As to the FPs, in general, E-FP first moves southward and then turns to the east, while W-FP first moves 
northward and then turns to the west, as indicated by the thick arrows in Figure~\ref{chp1029_HSIonMDI.ps}{\it b}.  
The evolution of the position of E-FP {\it relative} to W-FP 	
is shown in Figure~\ref{chp1029_HSIonMDI.ps}{\it c}.   
There is clearly a turning point 
which occurs at $t_1$=20:44~UT and divides the evolution of the FP positions into two phases: 
 (1) Phase~I (20:40:40--20:44:00~UT)
when the two FPs generally move toward each other in a direction essentially parallel to the NL, 
 (2) Phase~II (20:44:00--20:59:40~UT) 
when the FPs move away from each other mainly perpendicular to the NL.
According to \citet{Bogachev.HXR-motion.2005ApJ...630..561B}, this flare falls into their type~II
events during Phase~I and then type~III during Phase~II.
 Another signature of this two-phase division 
is the morphological	
transition at 20:43:20~UT, before which there are multiple FP sources,
but only two FPs present afterwards (see Fig.~\ref{chp1029_hsi_img_time.eps}). 
Below we describe in detail the HXR source evolution in the two phases.
%
%

We further decomposed the distance between the FPs into two components: perpendicular
and parallel to the NL as shown in Figure~\ref{chp1029_fp_motion.eps}{\it c}, where the two phases are 
divided by the vertical dotted line at $t_1$.  
As can be seen, the parallel distance ({\it asterisks}) first rapidly decreases at a velocity 
of $108 \pm 18 \, \km\ps$  given by the linear fit during Phase~I; 	
it then stays almost constant during Phase~II with a slow increase 	
($7 \pm 1 \, \km\ps$). In contrast, the perpendicular distance ({\it squares}) has a slow variation in Phase~I 	
($15 \pm 13 \, \km\ps$) and increases continuously at a velocity of $51 \pm 1 \, \km\ps$ in Phase~II. 
These velocities are comparable to those of the \trace EUV FPs found by 
\citet{SchrijverC.HudsonH.Oct28.2006ApJ...650.1184S} in the 2003 October 28 X17 flare
that occurred in the same active region as the flare under study.
From the speed ($\sim$60~km~$\ps$) and size ($<$1400~km) of EUV FPs, they inferred
the crossing time of the FP diameter or the excitation time scale of the HXR-producing 
electron beam in a single flare loop to be $\lesssim$23~s. 

Next we obtained the shear angle ($\theta$; Fig.~\ref{chp1029_HSIonMDI.ps}{\it c}) from the normal to the 
NL (parallel to the $y$-axis) to the W-to-E relative positional vector, which is shown as a function 
of time in Figure~\ref{chp1029_fp_motion.eps}{\it d}.   
This angle exhibits a fast decrease  (from $56\degree$ to $22\degree$)	
during Phase~I and a slow decrease (down to $12\degree$) 				
during Phase~II. 
An independent study by \citet{JiH2008ApJL.sigmoid}, with different identifications of the FPs in this flare
and thus larger scatter, also found a similar decrease of the shear angle 	
in two phases, which they referred to as {\it sigmoid} and {\it arcade} phases based on 
the X-ray morphology. 	
The apparent unshearing motions of the HXR FPs indicate that the later reconnected magnetic field lines
are less sheared. 
It can be seen	 
that \trace 195~\AA\ loops corresponding to the HXR FPs at early times (not shown) are indeed highly sheared.
Similar unshearing motions were observed in various wavelengths 	
\citep[e.g.,][]{ZirinTanaka.shear.1973SoPh...32..173Z,
MasudaS2001SoPh, SuYNGolubShear2007ApJ...655..606S}.
Note that an opposite process took place prior to the flare, that is,
strong photospheric shearing flows observed near the NL \citep[][as mentioned earlier]{YangG2004ApJ}. 
This process increased the shear of field lines and built up magnetic stress and 
free energy \citep{MetcalfT2005ApJ...623L..53M} in the system during the preflare phase. 

Finally we investigate the relationship between the FP and LT motions.
The division at $t_1$=20:44~UT between the two phases of the FP unshearing motions coincides (within 30~s) with 
the minimum of the estimated loop length (Fig.~\ref{chp1029_fp_motion.eps}{\it a})   
and the direction reversal of the apparent LT motion (Fig.~\ref{chp1029_fp_motion.eps}{\it b})  
noted above. Also the estimated LT height undergoes a general decrease during Phase I
when the HXR flux is on the rise.
A similar decrease of the LT altitude during the rising portion of the impulsive phase,
followed by a subsequent altitude increase, has been observed in many \hsi flares near the limb
\citep[e.g,][]{SuiL2003ApJ...596L.251S, SuiL2004ApJ...612..546S, LiuS2004ApJ...613L..81L, 
LiuW_2LT.2008ApJ...676..704L, HolmanG.LTstat.2005AGUFMSH13A0288H}.
For those flares a complete physical picture is obscured because the observed FP motions are
strongly subject to projection effects, but this drawback vanishes for the disk flare under study here.
Assuming that our semi-circular model loops yield reasonable estimates of the LT heights and loop lengths,
the above source motions, when taken together, suggest the following scenario:
(1) During Phase I, as the the reconnection site and thus the LT source migrate southward 
along the NL or the arcade, shorter and less sheared loops are energized, which leads
to the apparent decrease of the LT altitude and the fast unshearing motion of the FPs.
(2) During Phase II, as the the reconnection site migrates northward at gradually lower speeds,
longer and slightly less sheared loops are formed one above the other, and this results in 
the inferred increase of the altitude of the LT source and the separation of the FPs from the NL.

Phase~II is in agreement with the classical CHSKP picture of two-ribbon flares, while Phase~I is not.
One possible physical explanation for Phase~I%
 \footnote{	
 Considering the presence of multiple FPs (see Fig.~\ref{chp1029_hsi_img_time.eps}), 
 this phase might provide evidence for the tether-cutting model of 
 \citet{MooreR.tether-cutting.2001ApJ...552..833M}.}   
suggested by \citet{JiH.unshear2007ApJ...660..893J} is 
the magnetic ``implosion" conjecture \citep{HudsonH.implosion.2000ApJ...531L..75H}
that predicts contraction of field lines during a flare as a consequence of explosive energy release.
\citet{JiH.unshear2007ApJ...660..893J} further found that a magnetic field with shorter, lower lying, 
and less sheared field lines indeed contains less free energy.
Note that, in order to explain the LT descending motion, \citet{VeronigA2006A&A...446..675V} proposed 
a collapsing magnetic trap model, which, however, cannot explain the initial FP motion toward one another	
parallel to the NL.		
In contrast, the scenario of \citet{Somov.Bastille.2002ApJ...579..863S} for the approaching FPs
does not predict the decrease of the LT height.


\section{Temporal Correlations of Conjugate Footpoints}
\label{chp1029_corr}

We now examine the temporal evolution of and correlations between various quantities of the two conjugate FPs, 
particularly spectral, spatial, and magnetic field parameters, which are summarized in Table~\ref{chp1029_table_coef}.
%
\begin{table*}[bthp]	
\small  
\caption{Correlation coefficients and linear regressions between various parameters	
 of the conjugate footpoints for the full flare duration (20:40:40--20:59:40~UT). Magnetic field strengths are
 in units of 100 Gauss.}
\tabcolsep 0.06in	
\begin{tabular}{llcccll}
\tableline \tableline
 \multicolumn{2}{c} {Subscripts:}                     & $r_p$   &  \#~of     &  $r_s$   &  signif.  &  Linear regression (between  \\
 \multicolumn{2}{c} {\footnotesize 1: E-FP, 2: W-FP}  &         &  $\sigma$'s &        &   &  quantities in first two columns) \\
\tableline
$I_1$       & $I_2$         & 0.98  &  8   &  0.97   &  $10^{-35}$ & $I_2=(-0.3\pm 0.1)+(0.41 \pm 0.01) I_1$  \\
$\gamma_1$  & $\gamma_2$    & 0.90  &  7   &  0.89   &  $10^{-20}$ & $\gamma_2=(-0.5\pm 0.3)+(1.17 \pm 0.07) \gamma_1$  \\
$B_1$       & $B_2$        & 0.39  &  3   &  0.40   &  $10^{-3}$  & $B_2=(-9.2\pm 0.8)+(3.7 \pm 0.1) B_1$ \\   					
$B_1$    & $\log_{10} I_1$  & 0.50  &  4   &  0.49   &  $10^{-4}$  & $I_1= (0.012 \pm 0.003) \times 10^{(0.55 \pm 0.02) B_1}$  \\ 	
$B_2$    & $\log_{10} I_2$  & 0.82  &  6   &  0.84   &  $10^{-16}$ & $I_2= (0.19 \pm 0.01) \times 10^{(0.137 \pm 0.002) B_2}$  \\ 	
$\bar{B}$ & $\log_{10} \bar{I}$	 
                            & 0.77  &  6   &  0.84   &  $10^{-16}$ & $\bar{I}= (0.079 \pm 0.007) \times 10^{(0.252 \pm 0.005) \bar{B}}$  \\ 
$v_1$       & $v_2$         & 0.63  &  5   &  0.29   &  $10^{-2}$  & $v_2= (-93 \pm 10) + (2.8 \pm 0.1) v_1$\\

\tableline  \end{tabular}

\tablecomments{ 
 $r_p$: Pearson's linear correlation coefficient; \\
 \#~of $\sigma$'s: multiple of 1$\sigma$ uncertainty of $r_p$, where $\sigma=1/(57)^{1/2}=0.13$, 
    in which 57 is the number of data points (or time intervals); \\
 $r_s$ and signif.: Spearman's rank correlation coefficient and significance level;\\
 $I$ (photons~$\ps \, \pcms \, \pkeV$): HXR flux at 50~keV;\\
 $\gamma$: spectral index between 50 and 150~keV (regression done only for 20:40:40--20:52:40~UT);\\
 $B$ (in 100~G): magnetic field;\\
 $\bar{B}$ (in 100~G) and $\bar{I}$ (photons~$\ps \, \pcms \, \pkeV$): $B$ and $I$ averaged between E-FP and W-FP;\\
 $v$ (km~$\ps$): FP velocity.	
} 	
\label{chp1029_table_coef} \end{table*}
%
%

\subsection{Spectral Correlations}
\label{chp1029_corr_spec}

\begin{figure}[thbp]      
 \epsscale{1} 
 \plotone{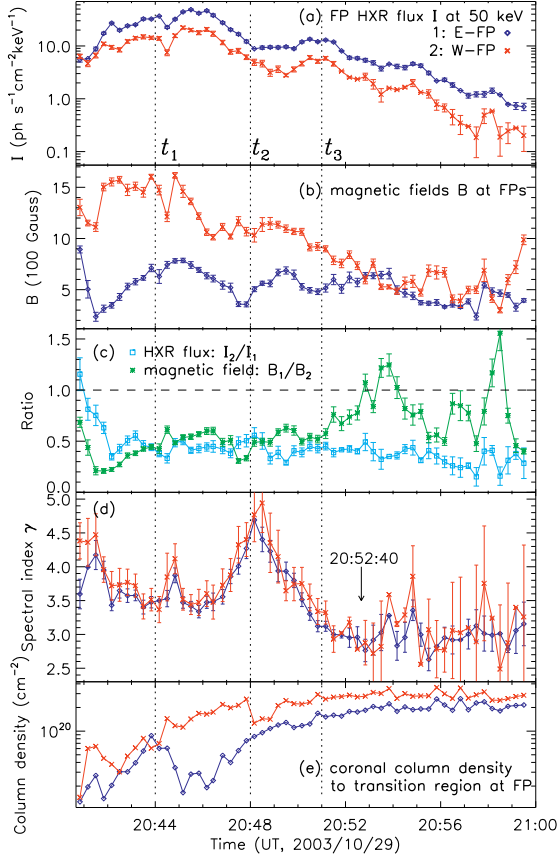}	
 \caption[History of the HXR fluxes of the two FPs and associated magnetic fields]
 {  
 Time profiles of X-ray and magnetic field parameters. 
   ({\it a}) HXR fluxes at 50~keV of E-FP ({\it blue diamonds}) and W-FP ({\it red crosses})
 obtained from power-law fits in the 50--150~keV range. 
 The same color and symbol convention for E- and W-FP holds for the other panels.	
 The vertical dotted lines correspond to the transitional times of $t_1$ and $t_2$ as shown in 
 Fig.~\ref{chp1029_lc.eps}, and $t_3$, the maximum of HXR Peak 2.
   ({\it b}) \soho MDI magnetic field strengths registered at the two FPs.
   ({\it c}) Ratios of the 50~keV fluxes (W-to-E) and magnetic fields (E-to-W)
 of the two FPs. 
   ({\it d}) HXR spectral indexes of the two FPs from the same fits as in ({\it a}).
 The arrow marks the end of the six 2 minute integration intervals for averaging the index values
 (see \S\ref{chp1029_corr_spec}).
    ({\it e}) Estimated coronal column densities from the edge of the LT source 
 to the	 
 transition region at the two FPs.
 }	\label{chp1029_vs_time.eps}
 \end{figure}

\begin{figure}[thbp]      
 \epsscale{1.15}		
 \plotone{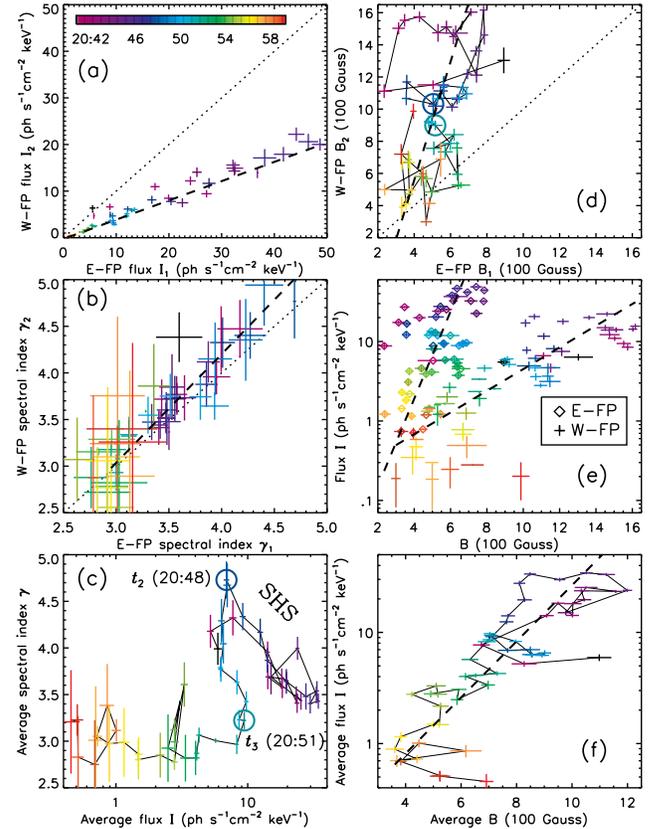}	
 \caption[Various correlations associated with the HXRs and magnetic fields of the two FPs.]
 { 
 Various correlations between the HXR and magnetic field parameters of the two FPs.
   ({\it a}) HXR flux at 50~keV (Fig.~\ref{chp1029_vs_time.eps}{\it a}) of W-FP vs.\ E-FP.
 The dashed line here and in other panels represents a linear regression of the data.
 The color bar indicates the common time evolution for all panels of this figure. 
   ({\it b}) Same as ({\it a}) but for the spectral indexes ($\gamma$) shown in Fig.~\ref{chp1029_vs_time.eps}{\it d}. 
 The linear regression is done for the data points up to 20:52:40~UT.
   ({\it c}) Average (between the two FPs) spectral index $\bar \gamma$ vs.\ average flux $\bar I$ 
 at 50~keV. 	  
 The two open circles here and in ({\it d}) mark the times $t_2$ and $t_3$ shown in Fig.~\ref{chp1029_vs_time.eps}.
 A ``soft-hard-soft" (SHS) variation is present early in the flare.
   ({\it d}) Same as ({\it a}) but for the magnetic field strengths of the two FPs as shown in 
 Fig.~\ref{chp1029_vs_time.eps}{\it b}.
   ({\it e}) Logarithmic HXR flux at 50~keV vs.\ the corresponding magnetic field strength for E-FP ({\it diamonds}) 
 and W-FP ({\it plus signs}). 
   ({\it f}) Same as ({\it e}) but for the average values.   
 }	\label{chp1029_multi_corr.eps}
 \end{figure}

Figure~\ref{chp1029_vs_time.eps}{\it a} shows the history of the photon fluxes of E-FP ($I_1$, {\it blue diamonds}) 
and W-FP ($I_2$, {\it red crosses}) at 50~keV obtained from the power-law fits in the 50--150~keV range mentioned
in \S~\ref{chp1029_obs_imgspec}.
We find that the two fluxes follow each other closely in their temporal trends
and E-FP is	always	
brighter than W-FP except for the first time interval.
The correlation of the fluxes can also be seen in 
Figure~\ref{chp1029_multi_corr.eps}{\it a} where one flux is plotted vs.\ the other. 
A linear regression 	
is shown as the thick dashed line and given in Table~\ref{chp1029_table_coef}.
The correlation coefficients listed in Table~\ref{chp1029_table_coef}
indicate a very high correlation in either a linear or a nonlinear sense. 
Such a correlation is expected for conjugate HXR FPs, since they are believed to be produced 
by similar populations of nonthermal electrons that escape the same acceleration region 
(believed to be at/near the LT source; \citealt{PetrosianV2004ApJ...610..550P};
\citealt{LiuW_2LT.2008ApJ...676..704L}) and travel down opposite legs of the same magnetic loop
to reach the chromosphere.

We show the corresponding power-law indexes ($\gamma$) of the two FPs vs.\ time in Figure~\ref{chp1029_vs_time.eps}{\it d}
and one index vs.\ the other in Figure~\ref{chp1029_multi_corr.eps}{\it b}.  
Again we find that the two indexes are closely correlated, as can be seen from the large correlation coefficients
(Table~\ref{chp1029_table_coef}).
The E-FP spectrum, however, is slightly harder than the W-FP spectrum,
which it is persistent most of the time. 	
The results from long integration intervals (2--3 minutes, not shown), which have better count statistics, 
exhibit the same pattern.
We averaged the index values of the six 2 minute intervals covering 20:40:40--20:52:40~UT,
after which the uncertainties become large due to low count rates.	
This average gives $\langle{\gamma_1} \rangle=3.63\pm 0.06$ for E-FP and $\langle{\gamma_2} \rangle =3.79 \pm 0.11$ 
for W-FP. Their difference of $\langle{\gamma_2} \rangle - \langle{\gamma_1} \rangle= 0.15 \pm 0.13$
is marginally significant at	
the 1$\sigma$ level.	

Let us compare the HXR fluxes and spectral indexes of the two FPs. As can be seen
in Figures~\ref{chp1029_vs_time.eps}{\it a} and \ref{chp1029_vs_time.eps}{\it d}, during HXR Peak~1
(before $t_2$=20:48~UT), the fluxes and indexes are anti-correlated, i.e., they follow the general	
``soft-hard-soft" (SHS) trend observed in many other flares 
\citep[e.g.,][]{ParksWinckler.SHS.1969ApJ...155L.117P, KaneAnderson.SHS.1970ApJ...162.1003K}.
However, during Peak~2 (after $t_2$),
the indexes decrease through the HXR maximum and then	
vary only slightly (with relatively large uncertainties) around a constant level of 3.0.
This trend can be characterized as	
``soft-hard-hard" (SHH). This flux-index relationship can also be seen in Figure~\ref{chp1029_multi_corr.eps}{\it c} where the 
index averaged between the two FPs is plotted against the average flux.
Note that the spectral index values during the late declining phase 	
of the flare are even smaller than those at the maximum of the main HXR Peak~1.  In this sense, 
the overall spectral variation can be characterized as ``soft-hard-soft-harder".  
As we noted earlier, there were energetic protons detected in 
interplanetary space by \goes and {\it ACE} following the flare.  These observations, 
when taken together, are consistent with the statistical result of  
\citet{KiplingerA_SEP-SHH_1995ApJ...453..973K} that this type of flare with progressive spectral hardening 
tends to be associated with SEP events \citep[also see][]{SaldanhaKrucker_SHH.SEP_2008ApJ}. 
As we also noted, strong	
gamma-ray line emission was detected during this flare 
\citep{HurfordG.OctNov.GR.2006ApJ...644L..93H}, which indicates a significantly large population of
accelerated protons at the Sun, but 	
the relation to the SEPs at 1~AU is unclear.

%
%
%
%

\subsection{Spatial Correlations}
\label{chp1029_corr_space}

We now examine the spatial evolution of two FPs.
In \S~\ref{chp1029_FPmotion} we focused on their relative motion, while here we 
compare their individual motions. 
Figure~\ref{chp1029_fp_motion.eps}{\it e} shows the perpendicular distance of each FP from the north-south NL 
({\it red dashed}, Fig.~\ref{chp1029_HSIonMDI.ps}{\it a}) as a function of time.
Linear fits of the full flare duration indicate mean velocities of 
$\langle v_\perp \rangle =36 \pm 1$ for E-FP and $13\pm 1\km \, \ps$ for W-FP. 
These velocities are similar to those found by \citet{XuY2004ApJ...607L.131X} for near infrared ribbons  
in this event.	
We also calculated the total velocities of the FP centroids,   
i.e., $v=(v_{\bot}^2+ v_{\|}^2)^{1/2}$,			
where $v_{\bot}$ and $v_{\|}$ are the components perpendicular and parallel 
to the NL, respectively.
The two resulting velocities 	
have a linear temporal correlation 
at a $5 \sigma$ level (see Table~\ref{chp1029_table_coef}), which again provides evidence of		
the causal connection between the conjugate FPs.
However, the individual component, $v_{\bot}$ or $v_{\|}$, alone does not exhibit any noticeable correlation between the two FPs.

  Figure~\ref{chp1029_fp_motion.eps}{\it a} shows the distances from the LT centroid to the centroids of
E-FP ($l_1$, {\it diamonds}) and W-FP ($l_2$, {\it crosses}) along the model semi-circular loop
(see, e.g., Fig.~\ref{chp1029_HSIonMDI.ps}{\it b}) as a function of time.
Each curve follows the same general increase as the corresponding distance from the NL shown
in Figure~\ref{chp1029_fp_motion.eps}{\it e}, but the distance to E-FP is smaller than that 
to W-FP most of the time.   
We estimated the coronal column densities from the LT source to the transition region at 
the two FPs (see Appendix \ref{chp1029_dens_factor} for details) to be 
$N_{{\rm tr}, \, i}= 0.5 n_{\rm LT} (l_i-r)$, where $i$=1, 2, 
using the distances $l_i$, LT density $n_{\rm LT}$ and equivalent radius $r$ 
obtained earlier (Fig.~\ref{chp1029_img_spec_time.eps}{\it c}). 	
The results in Figure~\ref{chp1029_vs_time.eps}{\it e} show that there is a large relative 
difference from 20:44 to 20:48~UT	
during HXR Peak~1 but a smaller difference during Peak~2.	
Implications of these different column densities will be addressed in \S~\ref{chp1029_asym_coldepth}
and \S~\ref{chp1029_asym_mirror_coldepth}.

\subsection{Magnetic Field Correlation}
\label{chp1029_corr_B}

The magnetic field strengths of the two FPs were obtained from {\it SOHO} MDI magnetograms 
(e.g., see Fig.~\ref{chp1029_HSIonMDI.ps}{\it a}) through the following steps:
(1) We first selected a preflare%
 \footnote{Note that magnetograms during the flare cannot be used due to temporary artificial changes in
  the measured field strength (see Appendix~\ref{chp1029_coalign}). 
  Permanent real changes have been observed before and after many X-class flares 
  \citep[e.g.,][]{WangH.Bchange.2002ApJ...576..497W, SudolJ.HarveyJ.Bchange.2005ApJ...635..647S},
  and have been interpreted as magnetic field changes in direction rather than in strength 
  \citep{SudolJ.HarveyJ.Bchange.2005ApJ...635..647S}.  Consequently we use the preflare 
  field as the best approximation available to us for the field during the flare.} 
magnetogram at 08:30:35~UT and coaligned it with the \hsi pointing and field of view 
using the offsets found in Appendix~\ref{chp1029_coalign}.
(2) For each time interval, the 90\% brightness contour (not necessarily a resolved source)
of each \hsi FP source was rotated back to the corresponding
position at the time of the MDI map to account for the solar rotation.  Then the MDI pixels enclosed in
this contour were averaged to give a value of the magnetic field for this FP, 
and the standard deviation of these pixels combined with the nominal 20~G MDI noise was used as the uncertainty.
(3) The above two steps were repeated for each of the ten MDI magnetograms recorded between 20:25 and 20:35~UT
at a one minute cadence (excluding the one at 20:28~UT that are contaminated by artificial pixel spikes).  
The average of the ten independent measurements was used as the final result for 
the magnetic field (Fig.~\ref{chp1029_vs_time.eps}{\it b}), and error propagation gave
the final uncertainty ($\sim$5--10\%).

As shown in Figure~\ref{chp1029_vs_time.eps}{\it b},	
the magnetic field strength of W-FP generally decreases with time, while that of E-FP fluctuates about
its mean value.  Most of the time (especially before $t_3$=20:51~UT), the W-FP field 	
is stronger than the E-FP field, 	
while their fractional difference generally decreases
as time proceeds. The temporal variations of the two field strengths 	
are only weakly correlated (again, particularly before $t_3$) 	
at the $3 \sigma$ level (see Table~\ref{chp1029_table_coef}), 
as can also be seen in Figure~\ref{chp1029_multi_corr.eps}{\it d}.


\subsection{Inter-correlations Among Spectral, Spatial, and Magnetic Field Parameters}
\label{chp1029_corr_inter}

Here we check the relationship between the HXR fluxes and the magnetic
fields of the two conjugate FPs.  We plot the logarithmic HXR flux vs.\ the magnetic field strength for each FP in 
Figure~\ref{chp1029_multi_corr.eps}{\it e}.  
As we can see, the flux is correlated with the field strength for each FP
(see Table~\ref{chp1029_table_coef} for the correlation coefficients and linear regressions). 
The logarithmic average flux ($\bar{I}$)	 
and magnetic field ($\bar{B}$) 
of the two FPs are shown one vs.\ the other in Figure~\ref{chp1029_multi_corr.eps}{\it f}.
A linear relationship, as shown by the thick dashed line with a correlation coefficient of $r_p=0.77\pm 0.13$, 
is clearly present. In other words, $\bar{I}$ is exponentially ({\it nonlinearly}) 
correlated with $\bar{B}$, the expression of which is listed in Table~\ref{chp1029_table_coef}.
Since $\bar{I}$ is correlated with $\bar{B}$, the ``soft-hard-soft"
type of relationship between $\bar{\gamma}$ and $\bar{I}$ during the
early phase (before $t_3$=20:51~UT) translates to that between $\bar{\gamma}$ and $\bar{B}$.
Namely, $\bar\gamma$ is anti-correlated with $\bar{B}$.

Finally we check the relationship of the apparent motions and magnetic fields of the two FPs.
As noted above, 
E-FP moves faster than W-FP 	
away from (perpendicular to) the magnetic NL, 
while E-FP is located in a weaker magnetic field.
This relationship	
means that, as expected, about the same amount of magnetic flux is annihilated from each polarity,
since $B \langle v_\perp \rangle$ is proportional to the magnetic reconnection rate.
However, the magnetic fluxes swept by the two FPs, $\int B \langle v_\perp \rangle dt$ integrated
over the full flare duration,  	
differ by 44\% of their average value.		
This is not surprising, 	
as \citet{FletcherHudson.ribbon2001SoPh..204...69F} found a similar flux mismatch	
and offered various explanations.   

\subsection{Discussion on Implications of Various Correlations}
\label{chp1029_interpret}

The above correlation (Fig.~\ref{chp1029_multi_corr.eps}{\it f}) between the average HXR flux 
and magnetic field strength  
reveals important information about the magnetic reconnection and particle acceleration processes.
Here we speculate on two alternative possible scenarios  	
in terms of contemporary acceleration theories.

1. The {\it nonlinear} (exponential) nature of the $\bar{I}$-$\bar{B}$ correlation 
suggests that particle acceleration is very sensitive to the magnetic field strength,
if we assume that the measured photospheric field strengths scale with
that in the coronal acceleration region.
The stochastic acceleration model of \citet{PetrosianV2004ApJ...610..550P} offers the
following predictions: 
(1) The level of turbulence that determines the {\it number} of accelerated electrons
is proportional to $\delta B^2$, where $\delta B$ is the magnetic field amplitude 
of plasma waves. 	
 The acceleration rate that determines the spectral {\it hardness} of accelerated electrons
is proportional to $\delta B^2/B$.	
An increasing field strength $B$ will result in an increasing flux and spectral
hardness of accelerated electrons, if both $\delta B^2$ and $\delta B^2/B$ also increase.
 (2) The relative efficiency of acceleration of electrons and thus their spectral {\it hardness}
increase with decreasing values of the ratio of electron plasma frequency to gyro-frequency, 
$\alpha \equiv \omega_{pe}/\Omega_e\propto 1/B$.
These predictions are qualitatively consistent with the observations that the magnetic field 
strength correlates with the HXR flux and anti-correlates with the spectral index.

2. Alternatively, noting the roughly constant velocities ($v_\perp$, Fig.~\ref{chp1029_fp_motion.eps}{\it e})
of the two FPs perpendicular to the magnetic NL, 	
the above $\bar{I}$-$\bar{B}$ correlation simply translates into 
the correlation between the HXR production rate and the magnetic flux annihilation rate or reconnection rate,
$Bv_\perp$. Furthermore, since $Bv_\perp$ is believed to be proportional 
to the electric field in the reconnection region \citep{ForbesLin.recon.2000JATP...62.1499F,
QiuJ2002ApJ...565.1335Q}, it then follows that the particle acceleration rate correlates with the electric field.  
According to the electric field acceleration model of \citet{HolmanG1985ApJ...293..584H} 
and \citet{BenkaS1994ApJ...435..469B}, 
a larger electric field results in a larger high-energy cutoff ($E_{\unit{max}}$) for the electron spectrum,
which can lead to a harder HXR spectrum \citep{HolmanGcutoff2003ApJ...586..606H}. This is 
consistent with the observed anti-correlation between the magnetic field strength and spectral index.
Note that in the classical \citep{PetschekH1964psf..conf..425P} model,
the small cross-section of the current sheet cannot	
account for the typically large flux of accelerated electrons	
of $10^{34}$--$10^{37}$ electrons $\ps$ 
\citep{MillerJ1997JGR...10214631M}, i.e., the so-called ``number problem".
However, observations and magnetohydrodynamic simulations \citep[e.g.,][]{KliemB.puls.recon.2000A&A...360..715K}
have implied that the reconnecting current sheet involves small-scale electric fields around multiple,
spatially separated magnetic X- and/or O-points in a fragmented topology
\citep[see a review by][]{AschwandenM2002SSRv..101....1A}. We speculate that, when particle acceleration 
takes place in such a fragmented current sheet, the number problem can be ameliorated 
\citep[also see a discussion by][]{HannahFletcher.test-particle.2006SoPh..236...59H}. 
In this case, the above discussion remains valid if the small-scale electric fields are
scaled with the macroscopic potential drop and thus with $Bv_\perp$. 
For comparison, we note that \citet{KruckerS2005AdSpR} studied the motion of E-FP {\it alone} in this flare and 
also found a rough temporal correlation between the HXR flux and reconnection rate, represented by 
$Bv$ or $B^2 v$, where $v$ includes the velocity both perpendicular and parallel to the NL. 

\section{Hard X-ray Footpoint Asymmetries}		
\label{chp1029_asym}

As mentioned above and partly noted by \citet{LiuW2004AAS}, \citet{XuY2004ApJ...607L.131X}, 
and \citet{KruckerS2005AdSpR}, the two conjugate FPs exhibit the following asymmetric characteristics:
 (1) the {\it brighter} E-FP is located in a {\it weaker}, negative magnetic field, 
while the {\it dimmer} W-FP is located in a {\it stronger}, positive field; 
 (2) the two FPs have very similar spectral shapes with E-FP being slightly harder; 	
 (3) E-FP is located {\it closer} to the LT 	
than W-FP;	
 (4) E-FP moves {\it faster} away from the magnetic NL than W-F.	
These asymmetries are summarized in Table \ref{chp1029_table_asym}.

We explore in this section different possibilities that can cause such asymmetries, particularly
the asymmetric HXR fluxes and spectra.  Various physical processes can contribute   
and they fall into two categories according to their origins: (1) asymmetry during particle acceleration,
and (2) asymmetry arising from 	
particle transport.  The second category includes effects
of magnetic mirroring and column density, which will be examined in what follows
(\S\ref{chp1029_asym_mirror}--\ref{chp1029_asym_coldepth}).
Other transport effects and the first category  
will be discussed later in \S~\ref{chp1029_asym_discuss}.
We use both the flux ratio $R_I \equiv I_2/I_1$ and the asymmetry 
\citep[c.f.,][]{Aschwanden.FPasym.1999ApJ...517..977A} defined by \citet{AlexanderD2002SoPh..210..323A}:
 \beq A \equiv (I_1 - I_2)/(I_1 + I_2) = (1-R_I) / (1+R_I), 
 \label{asymEq} \eeq 
to quantify the asymmetric HXR fluxes, with $A=\pm1$ being 100\% asymmetry and $A=0$ being perfect symmetry.
%
\begin{table}[thbp]	
\small	
\caption{Asymmetric characteristics of the conjugate footpoints (E-FP and W-FP): mean, median, 
and their E-to-W ratio of various quantities.}
\tabcolsep 0.04in	
\begin{tabular}{lrrrrrrr}
\tableline \tableline
                          & \multicolumn{3}{c} {Mean}   & & \multicolumn{3}{c} {Median} \\
                               \cline{2-4}                 \cline{6-8} \\
                             & E      &  W     & E/W    & &  E     &  W     & E/W \\
\tableline
$I$                          & 13.7   &  6.1   &  2.2   & &  8.9   &  4.6   &  1.9 \\
$B$                          & 520    &  960   &  0.55  & &  520   &  1010  &  0.51 \\
$\gamma$                     & 3.63   &  3.79  &  0.96  & &  3.4   &  3.5   &  0.96 \\
$l$                          & 41     &  60    &  0.69  & &  38    &  60    &  0.64 \\
$N_{\rm tr}$                 & 1.2    &  2.1   &  0.60  & &  1.2   &  2.5   &  0.50 \\
$v_\perp$                    & 36     &  13    &  2.8   & &   ---  &  ---   &  --- \\

\tableline  \end{tabular}

\tablecomments{
 $I$ (photons~$\ps \, \pcms \, \pkeV$): HXR flux at 50~keV; \\
 $B$ (Gauss): magnetic field strength; \\
 $\gamma$: spectral index between 50 and 150~keV;\\
 $l$ (arcsecs): distance from the LT centroid to the FP centroid along the semi-circular loop; \\
 $N_{\rm tr}$ ($10^{20}\pcms$): coronal column density from the edge of the LT source 
 to the transition region at the FP.\\
 $v_\perp$ ($\km \, \ps$) : velocity perpendicular to the NL; \\
 The mean spectral indexes are the averages of the results of the six 2 minute 
 integration intervals between 20:40:40 and 20:52:40~UT (see \S \ref{chp1029_corr_spec}), 	
 while all the other values listed here are from the results of the 57 short intervals throughout the flare.
 }
\label{chp1029_table_asym} \end{table}

\subsection{Magnetic Mirroring}
\label{chp1029_asym_mirror}

Asymmetric magnetic mirroring	
is commonly cited to explain asymmetric 
HXR fluxes observed at conjugate FPs.  We examine to what extent mirroring alone can explain the 
observations of this flare.  For simplicity, we make the following assumptions for our 	
analysis below:
 (1) Disregard all non-adiabatic effects of particle transport, 
i.e., energy losses and pitch-angle diffusion due to Coulomb collisions. 	
By this assumption, the magnetic moment of a particle is conserved and mirroring is the only
effect that changes the pitch angle when the particle travels in the loop and outside the acceleration region.  
 (2) Assume an isotropic pitch-angle distribution of the electrons at all energies upon release 	
from the acceleration region. 
 (3) Disregard details of bremsstrahlung, and assume that the nonthermal HXR flux is proportional to the 
precipitating electron flux at the FP.
 \footnote{This flux includes the precipitation of electrons previously reflected by mirroring back to 
 the acceleration region at the LT where they may be scattered and/or re-accelerated, presumably by turbulence.}

The loss-cone angle for magnetic mirroring is given as
$\theta_{i} = \arcsin (B_0/B_i)^{1/2}$, (i=1 for E-FP, 2 for W-FP),
where $B_0$ is the magnetic field strength at the injection site in the corona where particles escape 
from the acceleration region, and $B_i$ is the field strength at the $i$th FP in the chromosphere. 
By the isotropy assumption, the fractional flux of the forward moving electrons that will {\it directly} precipitate 
to the chromosphere (whose pitch angle is located inside the loss cone) can be evaluated by 
integrating over the solid angle \citep[also see][]{AlexanderD2002SoPh..210..323A}:
 \beq 
 F_i = {1 \over 2\pi} \int d\Omega = {1 \over 2\pi} \int_0^{2\pi} d\phi \int_0^{\theta_{i}} \sin\theta d\theta
    = 1-\mu_{i} \,,   
 \label{LCfluxEq} \eeq
where the pitch-angle cosine is $ \mu_{i}= \cos \theta_{i} = (1-B_0/B_i)^{1/2}$. 
If there is strong mirroring, i.e., $B_0 \ll B_i$, we have $\mu_{i} \simeq 1- B_0 /(2 B_i)$, 
and if there is no mirroring, i.e., $ B_0=B_i$, we have $\mu_{i}=1$.
By our assumptions (1) and (3) above, such a fraction should be independent of electron energy and is 
proportional to the HXR flux $I_i$ at the corresponding FP. It then follows that
 \begin{eqnarray}
 R_I  & \equiv  & {I_2 \over I_1} = {F_2 \over F_1} = { 1- \mu_2 \over 1 - \mu_1 } 	
 \\
 & \simeq &
 \cases{
         {B_1 \over B_2}  \equiv R_B^{-1}, & if $B_0 \ll B_i, \, (i=1, \, 2)$, \cr
         {B_1 \over 2 B_2}  \equiv R_B^{-1} / 2, & if $B_0=B_1, \, B_0 \ll B_2$, 
       }
 \label{LC_flux_ratio_Eq} 
 \end{eqnarray}
the second case of which corresponds to the possibility that mirroring occurs only at one FP,
but the required condition $B_1 \ll B_2$ does not apply to this flare.
In either case, the HXR flux ratio should be correlated with the inverse of the magnetic field ratio, $R_B^{-1}$.
This result is consistent with that of the strong diffusion case obtained by \citet{MelroseDWhiteS1981JGR.86.2183Asym}.

As shown in Table~\ref{chp1029_table_asym}, the mean/median HXR flux of E-FP is about twice that of W-FP,
while the mean/median magnetic field strength of E-FP is about a factor of two smaller.  This is consistent with
the mirroring effect in the average sense of the whole flare duration.  
We can check if this relationship also holds at different times.
Figure~\ref{chp1029_vs_time.eps}{\it c} shows the W-to-E ratio ($R_I= I_2/I_1$) of the HXR fluxes and
the E-to-W ratio ($R_B^{-1}= B_1/B_2$) of the field strengths of the two FPs as a function of time. 
We find a temporal correlation between the 
two ratios during the first $\sim$3 minutes when both first decrease and then increase. 
In the middle stage (20:43--20:51~UT) of the flare, both ratios remain roughly constant with marginal fluctuations 
and similar mean values of $\langle{R_I}\rangle=0.43$ and $\langle{R_B^{-1}}\rangle= 0.50$.	
After 20:51~UT, the magnetic field ratio increases significantly with large
fluctuations and exceeds unity in 6 of the 57 time intervals,	
while the flux ratio remains at about the same level as before.
The behavior of the two ratios before 20:51~UT is expected from magnetic mirroring, but their
significant difference after 20:51~UT cannot be explained by mirroring alone.  


\subsection{Column Density}
\label{chp1029_asym_coldepth}

Another transport effect that can cause asymmetric HXR FPs is different coronal column densities experienced by electrons
in traveling from the acceleration region to the transition region at the two FPs (\citealt{Emslie2003ApJ0723}; 
\citealt[][p.~73]{LiuW2006PhDT........35L}).  The effective column density is
$N_{\rm tr, \, eff}= N_{\rm tr}/ \langle\mu\rangle$, where $\langle\mu\rangle$ is the average pitch angle
cosine, and $N_{\rm tr}= \int_0^{s_{\rm tr}} n(s) ds$ is the coronal column density to the transition region at
distance $s=s_{\rm tr}$ along the magnetic field line with $n(s)$ being the ambient electron number density.
A difference in $\langle\mu\rangle$, $s_{\rm tr}$, and/or $n(s)$ between the two legs of the flare loop
can lead to different effective column densities.
 (1) Different pitch-angle distributions can be caused by asymmetric magnetic mirroring and/or asymmetric
acceleration.
 (2) Different path lengths $s_{\rm tr}$ can be caused by  
a magnetic reconnection site located away from the middle of the loop \citep{FalewiczR2007A&A}.	
 (3) Different densities $n(s)$ can also occur	
because magnetic reconnection takes place between field lines that 
are previously not connected and their associated densities	
are not necessarily the same. It takes time (on the order of the sound travel time,	
$\gtrsim$ tens of seconds) for the newly reconnected loop to reach a density equilibrium, 
but the observed HXRs could be produced before then.  

Column density asymmetry affects the FP asymmetry in two ways, since both energy losses and pitch-angle scattering
due to Coulomb collisions take place at about the same rate that is proportional to the column density:
  (1) Column density asymmetry is related to energy losses  
and the way we calculate the FP photon flux in \S~\ref{chp1029_obs_imgspec} 
(integrating HXR photons primarily produced below the transition region). 
Electrons with an initial energy of $E$ are stopped after traveling through a column density 
$N_{\rm stop} (\pcms) / \approx 10^{17} [E(\keV)]^2$. If $N_{\rm stop}$
is smaller than or comparable to the column density to the transition region ($N_{\rm tr}$), 
in one half of the loop with a {\it larger} column density,	
there are {\it more} electrons stopped in the leg and thus less electrons reaching
the transition region. This results in {\it more} HXRs produced in the leg and less HXRs beneath the 
transition region (counted as the FP flux).
%
  (2) Different Coulomb scattering rates result from different column densities on the two sides
of the loop, which can cause different pitch-angle distributions, even if the particles are injected 
with symmetrical pitch angles	
from the acceleration region. 
%
%
%
 \begin{figure*}[thbp]        
 \epsscale{1}	
 \plotone{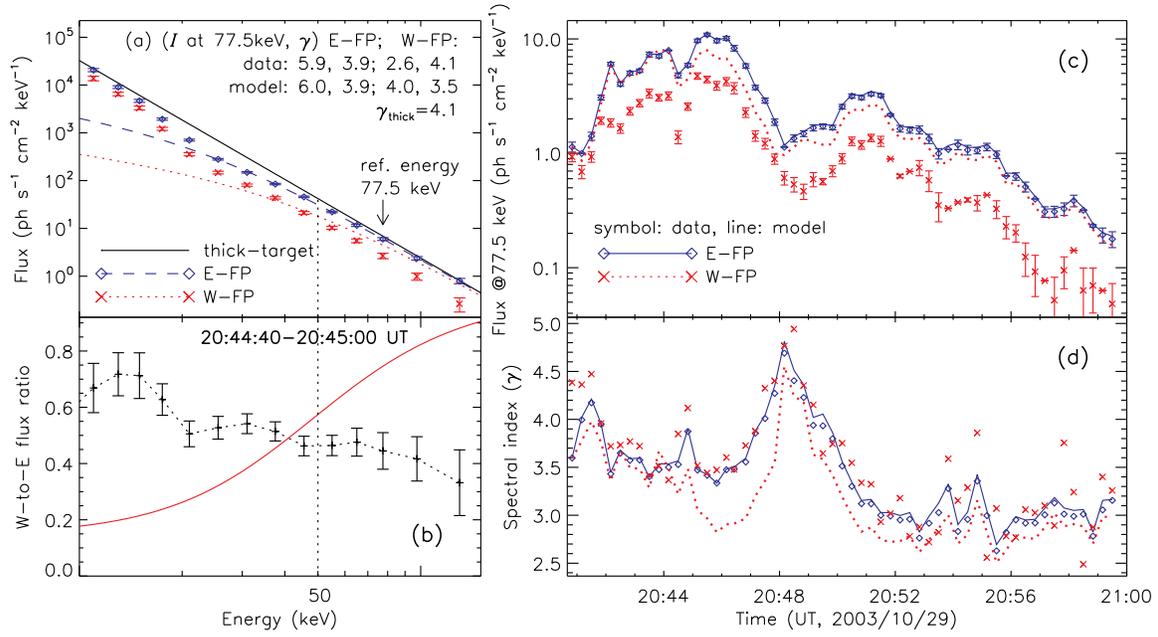}	
 \caption[Effects of column densities on FP asymmetry]
 {
 Effects of asymmetric coronal column densities.
   ({\it a}) Photon fluxes of E-FP ({\it diamonds}) and W-FP ({\it crosses}) vs.~energy at 20:44:40--20:45:00~UT
 (same as Fig.~\ref{chp1029_img_spec.eps}{\it a}), superimposed with model fluxes ({\it lines}) evaluated below 
 the transition region 	
 using eq.~(\ref{Ik_tau_Eq1}).
 The solid line shows the power-law thick-target flux with an index of $\gamma_{\rm thick}=4.1$. 
 The legend shows the fluxes ($I$) at a reference energy of 77.5~keV
 and the spectral indexes ($\gamma$) above 50~keV of the two FPs for both the data 
 and model. 
   ({\it b}) Ratio of the W-to-E FP fluxes shown in ({\it a}), with the plus signs for the data
 and the solid line for the model (given by eq.~\ref{RI_tau_Eq}).
   ({\it c}) Observed ({\it symbols}) and modeled ({\it lines}) fluxes of E- and W-FP at 77.5~keV vs.~time 
(cf. Fig.~\ref{chp1029_vs_time.eps}{\it a} at 50~keV).	
   ({\it d}) Same as ({\it c}) but for the spectral indexes above 50~keV. The data is the same as in 
 Fig.~\ref{chp1029_vs_time.eps}{\it d} but without error bars for clarity.
 } \label{chp1029_model_coldep_asy.eps}
 \end{figure*}
%

Focusing on the energy dependence of FP HXR asymmetry, we present below an estimate 	
of the column density effect {\it alone}, while assuming no magnetic mirroring 	
and identical pitch-angle distributions (same $\langle\mu\rangle$) in the two loop legs.
The relevant formulae	
are derived in Appendix \ref{chp1029_coldep_math}. 
We further assumed that identical power-law electron fluxes with a spectral index $\delta$ 	
are injected into the two legs of the flare loop, which have the same ambient
density but different path lengths $s_{\rm tr}$ to the FPs.
Our goal is to examine if this scenario can yield 	
photon fluxes and spectra consistent with observations for both FPs. 
For each of the 57 time intervals shown in Figure~\ref{chp1029_vs_time.eps}, 
we first used the E-FP column density $N_{\it tr}$ (see \S\ref{chp1029_corr_space} and
Fig.~\ref{chp1029_vs_time.eps}{\it e}) from the edge of the LT source (assumed to be the acceleration region)
to obtain its dimensionless form $\tau_{\rm tr} = N_{\it tr} / (5\E{22} \pcms)$, which ranges from
$5 \E{-4}$ to $1 \E{-2}$. We then substituted $\tau_{\rm tr}$ into
  \beq
    I_\unit{FP}(\tau_{\rm tr}, k) 
    = A_0 k^{-(\delta-1)} \left(1 + \tau_{\rm tr} \frac{k+1}{0.37 k^2}\right)^{1- \delta/2} \,,
  \label{Ik_tau_Eq1} \eeq
rewritten from equation~(\ref{Ik_tau_Eq}), where $A_0$ is the normalization for the thick-target flux and 
$k$ is the photon energy in units of the rest electron energy 511~keV.	
With this equation, we fitted the E-FP spectrum above 50~keV in a least-squares sense by
iteratively adjusting the free parameters $A_0$ and $\delta$. 
Using the resulting $A_0$ and $\delta$ and W-FP's $\tau_{\rm tr}$, we then calculated the W-FP spectrum 
by equation~(\ref{Ik_tau_Eq1}) 	
and the W-to-E flux ratio by equation~(\ref{RI_tau_Eq}).

Figure~\ref{chp1029_model_coldep_asy.eps}{\it a} shows an example of the spectra of the two FPs and
their model predictions, together with the corresponding thick-target spectrum produced 
by the same power-law electron flux.
We only trust the observed FP spectra $>$50keV due to pileup, as noted earlier.
As expected, the model FP fluxes are reduced from the thick-target flux, especially at low energies,
because low-energy electrons are more susceptible to collisional energy loss and pitch-angle scattering.
This results in a	
spectral flattening (hardening)
in the FP 	
X-ray spectrum. Because of its larger column density, the W-FP's model spectrum exhibits more flux reduction 
at a given energy and a flattening to a higher energy.
Above 50~keV the model spectrum of the brighter E-FP fits the data very well.	
However, that of the dimmer W-FP does not fit the data at all, since the model flux is
much greater (e.g., at 77.5~keV $I=4.0$ vs.~2.6 photons~$\ps \, \pcms \, \pkeV$)
and harder  ($\gamma=3.5$ vs.~4.1) than the observed flux, and even harder than the E-FP flux
($\gamma=3.5$ vs.~3.9). 
This can be best seen in Figure~\ref{chp1029_model_coldep_asy.eps}{\it b} that
shows the data ({\it plus signs}) and model ({\it solid line}) ratios of the W-to-E FP flux.
The data ratio generally decreases with energy or stays roughly constant above 50keV within uncertainties, 
but the model ratio is an increasing function of energy.
These trends generally hold throughout the flare as can be seen from the history of HXR fluxes and spectral indexes
shown in Figures~\ref{chp1029_model_coldep_asy.eps}{\it c} and \ref{chp1029_model_coldep_asy.eps}{\it d}.

In summary, the model predicts a much harder photon spectrum for the dimmer W-FP with the larger coronal column density, 
while according to the observations the dimmer W-FP is as hard as or
slightly softer than the brighter E-FP (see Fig.~\ref{chp1029_model_coldep_asy.eps}{\it d}). 
\citet{Saint-Hilaire2008SolPh.FPasym} reported similar results that the majority of the brighter 
FPs in their 172 pairs of FPs during the HXR peaks of 53 flares tend to have harder spectra. 
 In addition, the differences between the model HXR
fluxes of the two FPs are too small to explain the observations (Fig.~\ref{chp1029_model_coldep_asy.eps}{\it c}).
One may attempt to increase the difference between the column densities in order to increase the flux difference
and thus to merge this gap between the model and data, but the discrepancy of the spectral 
indexes would be exacerbated since the W-FP spectrum would be relatively even harder.
Therefore, we conclude that the column density effect alone cannot provide a self-consistent explanation 
for both the HXR fluxes and spectra of the asymmetric FPs observed here.
%

\citet{FalewiczR2007A&A}, however,	
found in three flares that the HXR flux ratios    
of asymmetric FPs were consistent (within a factor of 2) 	
with the predictions from asymmetric column densities. While that scenario may apply to those flares, 
we should note that their broad band {\it Yohkoh} observations set less stringent constraints 
than our high resolution \hsi observations, which can lead to different conclusions.
In particular, their analyses were limited to images in the M1 (23--33~keV) and M2 (33--55~keV) bands,
where the column density effect is more pronounced and contaminated from thermal emission is
possible, while our observations cover a higher and wider energy range of 50--150~keV.

\subsection{Magnetic Mirroring and Column Density Combined}
\label{chp1029_asym_mirror_coldepth}

We have seen from the above discussions that each of the two transport effects {\it alone} can only explain to some
extent the observed FP asymmetries:  (1) Asymmetric magnetic mirroring is consistent with the asymmetric HXR
fluxes in the average sense of the flare duration, but it has difficulties in accounting for the flux asymmetry
later in the flare.
(2) Asymmetric column densities in the two legs of the flare loop are qualitatively consistent with the
asymmetric HXR fluxes,	
but their quantitative predictions of fluxes and spectral hardness contradict the observations.
These two transport effects, in reality, operate at the {\it same time}, because electrons experience Coulomb
collisions while being mirrored back and forth in the loop, and thus the collisionless (adiabatic) assumption 
that we adopted earlier for simplicity for magnetic mirroring needs to be dropped.
In particular, since W-FP has stronger mirroring (than E-FP), the average pitch angle of electrons impinging there
is larger, and thus the effective column density $N_{\rm tr, \, eff}= N_{\rm tr}/ \langle\mu\rangle$ is
greater than previously thought.  
 This can enhance the column density asymmetry.
In what follows, we attempt to provide an explanation for some aspects of the observations by
combining the two effects.

From the above discussion and the observations presented in \S~\ref{chp1029_corr},
we should pay attention to the distinction between the two HXR peaks.
As shown in Figure~\ref{chp1029_vs_time.eps}{\it c}, during Peak~1 ($< t_2$=20:48~UT) the FP HXR flux asymmetry 
seems to be mainly controlled by magnetic mirroring,
while during Peak~2 ($>t_2$), 
especially after the HXR maximum at $t_3$=20:51~UT, 
this control seemingly fails. 
A viable explanation for the two-peak distinction is that (1) at early times, the densities 
(Fig.~\ref{chp1029_img_spec_time.eps}{\it d}) and lengths (Fig.~\ref{chp1029_fp_motion.eps}{\it a}) 		
of the loops are small, resulting in small coronal column densities (Fig.~\ref{chp1029_vs_time.eps}{\it e}) 
from the acceleration site to the FPs.	
Energy losses and pitch-angle scattering 
due to Coulomb collisions are less important, and therefore the rates of electron precipitation to the FPs 
are mainly governed by mirroring.
(2) Later in the flare,
as the loop densities and lengths have increased considerably, the column densities become larger and the collisional effects 
become more important than before in shaping the observed FP flux asymmetry.  
In addition, since magnetic mirroring depends on 
the gradient $d\ln B/dN$ \citep[e.g.,][]{LeachJ1981ApJ...251..781L}, this effect becomes less important when 
the column density $N$ increases faster than the relative change of magnetic field from the LT to the FP, 
which is possibly the case later in the flare.
Therefore, at later times, the prediction of magnetic mirroring alone tends to deviate from the data,
which might be explained by the two transport effects combined.  The outcome of the combination 
can be modeled quantitatively as discussed later in \S~\ref{chp1029_conclude} for future work.

There are several coincidences with the two-peak division which seem to have causal connections:
 (1) As noted in \S~\ref{chp1029_corr_spec}, the correlation between the HXR fluxes and
spectral indexes (Figs.~\ref{chp1029_vs_time.eps}{\it a} and \ref{chp1029_vs_time.eps}{\it d}) can be described as
common ``soft-hard-soft" during Peak~1 and as ``soft-hard-hard" during Peak~2.
 The spectral hardening	at later times	
may be associated with the increasing loop column densities (Fig.~\ref{chp1029_vs_time.eps}{\it e}),
due to collision-caused hardening mentioned above 
(\S\ref{chp1029_asym_coldepth}). 	
 (2) During Peak~1, the magnetic fields at the two FPs are weakly correlated with each other 
(Fig.~\ref{chp1029_vs_time.eps}{\it b}),
while this correlation becomes progressively nonexistent during Peak~2, especially after its maximum $t_3$=20:51~UT, 
possibly because of longer loops.
 (3) 
The transition ($t_2$=20:48~UT) between the two HXR peaks coincides with the sudden jump in the positions
of both FPs (Figs.~\ref{chp1029_HSIonMDI.ps}{\it a} and \ref{chp1029_fp_motion.eps}{\it e}), 
the dip in the loop length (Fig.~\ref{chp1029_fp_motion.eps}{\it a}), 	
and the valley in the magnetic field strengths	
(Fig.~\ref{chp1029_vs_time.eps}{\it b}).
This points to the start of the new episode of energy release of Peak~2, presumably associated with a new series 
of loops that have physical conditions different from those in Peak~1. This transition may be related
to the different behaviors of magnetic mirroring during the two peaks noted above.


\subsection{Discussion on Other Asymmetry-causing Effects}	
\label{chp1029_asym_discuss}


Here we briefly discuss transport effects other than magnetic mirroring 
and column density, and acceleration related effects that can contribute to 	
the observed HXR flux and spectral asymmetries.		

Asymmetry during the {\it transport process} ---
(1) {\bf Non-uniform target ionization}: 
In the above analysis we assumed a fully ionized background
in the path of high-energy electrons, but in reality the background targets vary from fully ionized in the corona 
to neutral in the chromosphere. The presence of neutral atoms reduces the rates of long-range collisional
energy losses and pitch-angle scattering and thus increases the bremsstrahlung efficiency 
\citep{BrownJ.ioniz.1973SoPh...28..151B, LeachJ1981ApJ...251..781L, KontarE.ioniz.2002SoPh..210..419K}.
In this flare, at the E-FP with weaker magnetic mirroring electrons can penetrate deeper into the 
chromosphere and thus encounter more neutral atoms.	
This may result in a higher HXR flux and harder spectrum in the 50--150~keV 
range at E-FP than at W-FP, qualitatively consistent with the observations.
(2) {\bf Relativistic beaming and photospheric albedo}: At the E-FP with weaker mirroring, the angular distribution of electrons are more concentrated
to the forward direction down to the photosphere. When the FPs are seen on the solar disk from above,
the increasing importance with energy of the forward relativistic beaming effect 
\citep{McTiernanJ.c2limb.1991ApJ...379..381M} results in a smaller fraction of high-energy photons 
emitted upward at E-FP. Meanwhile, since relatively more photons are (beamed) emitted downward at E-FP, albedo
or Compton back-scattering \citep{Langer.Petrosian1977ApJ...215..666L, BaiT.albedo1978ApJ...219..705B}
is stronger there. Both effects can cause a softer X-ray spectrum at E-FP than at W-FP,
competing with other effects mentioned above, which may explain why the spectral indexes are so close
($\langle{\gamma_2} \rangle - \langle{\gamma_1} \rangle= 0.15 \pm 0.13$).	
(3) {\bf Return currents} and the associated electric field decrease the energy 
of the downward-streaming electrons, with the major impact being on the lower-energy electrons 
\citep{ZharkovaV.return-current.2006ApJ...651..553Z}.
Different precipitating electron beam fluxes in the two legs of the flare loop may induce different 
return current densities, and thus can result in different HXR fluxes and spectral shapes at the two FPs. 



Intrinsic asymmetry arising from the particle {\it acceleration process} ---
An energy-dependent FP HXR flux asymmetry ($|A|$, eq.~[\ref{asymEq}]),
which has a maximum in the intermediate energy range (20--40~keV) and decreases toward both
low and high energies, was found by \citet{AlexanderD2002SoPh..210..323A}.
This was attributed to an asymmetric, energy-dependent accelerator from which more electrons 
are injected preferentially into one of the two legs of the loop \citep{McClementsK2005ApJ}.
Asymmetric electron beams 	
can be produced by the electric field in a reconnection current sheet \citep{ZharkovaV.asym-acc.2004ApJ...604..884Z}.
For the flare under study no reliable asymmetry can be obtained below 50~keV due to pulse pileup;
above 50~keV asymmetry either increases with energy (see the flux ratio in Fig.~\ref{chp1029_img_spec.eps}{\it b}) 
or remains constant, opposite to the decreasing asymmetry reported by \citet{AlexanderD2002SoPh..210..323A}
in this energy range.  From {\it Yohkoh} HXT data \citet{Aschwanden.FPasym.1999ApJ...517..977A} also 
found no general energy-dependent pattern of flux asymmetry. Furthermore, 
if acceleration by plasma waves is the dominant mechanism, 
it is difficult to realize an asymmetric particle accelerator in 
the turbulence region due to frequent scatterings. Whether the scenario proposed by \citet{McClementsK2005ApJ} 
is the rule or an exception thus remains an open question.


\section{Summary and Conclusion}	
\label{chp1029_conclude}

We have presented imaging and spectral analysis of the \hsi observations of the 2003 October 29 
X10 flare showing two conjugate HXR footpoints (FPs), which are well-defined during the bulk of
the flare duration. One FP lies to the east (E-FP) and the other to the west (W-FP) of 
the north-south magnetic neutral line (NL).
This flare provides a unique opportunity to study in great detail the spatial, temporal, and 
spectral properties of the FPs and their associated magnetic fields. The impulsive phase was
relatively long ($\sim$20 minutes), HXR fluxes were detected by \hsi at energies up to hundreds of keV, 
and it was located close to disk center, resulting in minimum projection effects and
excellent magnetic field measurements from \soho MDI. 
Our main findings regarding the unshearing motions, various correlations, and asymmetric characteristics
of the two FPs are as follows.


1. {\bf Two-phase FP unshearing and loop-top (LT) motions} are observed in this flare.
In Phase I the two identified FPs become closer to each other 
as they rapidly move almost anti-parallel to the magnetic NL, while in Phase II they move away from 
each other slowly, mainly perpendicular to the NL (Fig.~\ref{chp1029_HSIonMDI.ps}{\it a}). 
 In other words, the shear angle $\theta$ between the normal to the NL and 
the line connecting the two FPs exhibits a fast and then slow decrease
from $56\degree$ to $12\degree$ (Fig.~\ref{chp1029_fp_motion.eps}{\it d}).
This suggests that later reconnected magnetic field lines 		
are less sheared (closer to a potential field), which is consistent with early
observations in HXRs and other wavelengths \citep[e.g.,][]{ZirinTanaka.shear.1973SoPh...32..173Z}.
 More importantly, the transition between the two phases 
coincides with the direction reversal of the apparent motion of the LT source along the NL 
(Fig.~\ref{chp1029_fp_motion.eps}{\it b}), 
and the minima of the estimated loop length (Fig.~\ref{chp1029_fp_motion.eps}{\it a}) 
and LT height (Fig.~\ref{chp1029_fp_motion.eps}{\it b}).  
This suggests that the {\it initial decrease} of the LT altitude observed in many other
\hsi flares \citep[e.g.,][]{SuiL2003ApJ...596L.251S}
may be associated with shorter loops during the {\it fast unshearing} motion phase when
the reconnection site propagates along the arcade. 		
A possible explanation for this early phase is the implosion conjecture 
\citep{HudsonH.implosion.2000ApJ...531L..75H} that predicts contraction of field lines
during a solar explosion including flares and CMEs.

2. There are {\bf correlations} among the temporal evolutions of various quantities (Table~\ref{chp1029_table_coef}),
some of which exhibit distinctions between the two HXR peaks
(division at 20:48~UT):	 
({\it a})	
  The HXR fluxes (Figs.~\ref{chp1029_vs_time.eps}{\it a} and \ref{chp1029_multi_corr.eps}{\it a}) and 
spectral indexes (Figs.~\ref{chp1029_vs_time.eps}{\it c} and \ref{chp1029_multi_corr.eps}{\it b}) 
of the two FPs are strongly correlated.
 This is evidence that the two HXR sources are from conjugate FPs at the two ends of the same magnetic loop.
({\it b})	
  The HXR flux and spectral index of each FP show a commonly observed ``soft-hard-soft" evolution   
(Figs.~\ref{chp1029_vs_time.eps}{\it a}, \ref{chp1029_vs_time.eps}{\it d}, and \ref{chp1029_multi_corr.eps}{\it c})
during HXR Peak~1, while during Peak~2 
the evolution becomes ``soft-hard-hard".
({\it c})	
  The magnetic field strengths at the two FPs also  
exhibit some temporal correlation 
(Figs.~\ref{chp1029_vs_time.eps}{\it b} and \ref{chp1029_multi_corr.eps}{\it d}) particularly during Peak~1,
consistent with the conjugate FPs identification.
({\it d})	
  The FP HXR fluxes exponentially		 
correlates with the magnetic field strengths (Figs.~\ref{chp1029_multi_corr.eps}{\it e} and \ref{chp1029_multi_corr.eps}{\it f}),
which also anti-correlates with the spectral indexes during Peak~1.
 These correlations suggest that stronger		
magnetic fields, and/or larger reconnection rates or 	
larger electric fields in the reconnection region are responsible for producing larger fluxes and harder spectra for
the accelerated electrons and thus the resulting HXRs. This is in qualitative agreement with the predictions
of the stochastic acceleration model \citep{PetrosianV2004ApJ...610..550P} and the electric field acceleration
model \citep{HolmanG1985ApJ...293..584H}.

3. Various {\bf asymmetries} are observed between the conjugate FPs (Table \ref{chp1029_table_asym}):
({\it a}) 
On average,
the eastern footpoint (E-FP) HXR flux is 2.2 times higher than that of the western footpoint (W-FP;
Fig.~\ref{chp1029_vs_time.eps}{\it a}), while its magnetic field strength is 1.8	
times weaker (520~G vs.~960~G; Fig.~\ref{chp1029_vs_time.eps}{\it b}).
This is consistent with asymmetric magnetic mirroring (\S\ref{chp1029_asym_mirror}).
({\it b})	
The average estimated coronal column density from the edge of the LT source 
(assumed to be the acceleration region) to the transition region at E-FP is 1.7 times smaller than	
that of W-FP ($1.2 \E{20}$ vs.~$2.1 \E{20} \pcms$; Fig.~\ref{chp1029_vs_time.eps}{\it e}). 
This qualitatively agrees with the HXR flux asymmetry, because a larger coronal column density results 
in more HXRs produced in the loop legs and thus less HXRs emitted from the FP below the transition region,
especially at low energies (Fig.~\ref{chp1029_model_coldep_asy.eps}; \S\ref{chp1029_asym_coldepth}).
({\it c})	
The photon spectra above 50~keV of the two FPs are almost parallel to each other (Fig.~\ref{chp1029_img_spec.eps}{\it a}),
with the brighter E-FP being consistently slightly harder than the dimmer W-FP (Fig.~\ref{chp1029_vs_time.eps}{\it d}). 
Their mean index values $\langle{\gamma_1}\rangle= 3.63\pm 0.06$ and $\langle{\gamma_2}\rangle = 3.79\pm 0.11$ 
have a marginally significant difference of $\langle{\gamma_2} \rangle - \langle{\gamma_1} \rangle= 0.15 \pm 0.13$.
In other words, the W-to-E ratio of the photon fluxes is a constant or a slightly decreasing 
function of energy. This	
contradicts the column density effect which would produce a harder spectrum at 
the dimmer W-FP (Fig.~\ref{chp1029_model_coldep_asy.eps}).  
({\it d})	
As expected from asymmetric magnetic mirroring, there is a temporal correlation between the W-to-E HXR flux ratio
and the E-to-W magnetic field ratio. However, this correlation only holds during HXR Peak~1 but gradually 
breaks down during Peak~2 (Figs.~\ref{chp1029_vs_time.eps}{\it c}).  
We suggest that a combination of the asymmetric magnetic mirroring and column density effects
could explain this variation (\S\ref{chp1029_asym_mirror_coldepth}). 
Specifically, since the column densities in later formed loops are larger (Fig.~\ref{chp1029_vs_time.eps}{\it e}), 
collisions are more important at later times, making the HXR flux ratio deviate from the prediction of mirroring alone.



In our analysis we have treated the magnetic mirroring and column density effects separately in order 
to make the problem analytically tractable, yet without loss of the essential physics.  
However, in reality, the two effects are coupled and they should be studied together self-consistently to
obtain a quantitative model prediction.	
This is done with the Fokker-Planck particle transport model of \citet{LeachJ1981ApJ...251..781L} 
in a converging magnetic field geometry. Results from such an analysis will be presented in a future publication.
In addition to numerical modeling, we have started a statistical study of \hsi flares showing
double FP sources that are close to disk center and thus have less projection effects.	
We hope to conduct future joint observations with {\it RHESSI}, {\it Hinode},
and the {\it Solar Dynamic Observatory} to obtain  
more advanced measurements of the magnetic fields at FPs.	
These future investigations will help improve our understanding of the underlying physics of asymmetric HXR FPs.



\acknowledgements
W. Liu was supported by an appointment to the NASA Postdoctoral Program at Goddard Space Flight Center, 
administered by Oak Ridge Associated Universities through a contract with NASA. 
This work was also supported in part by NSF grant ATM-0648750 at Stanford University. 
Work performed by B. Dennis and G. Holman was supported in part by 
the NASA Heliophysics Guest Investigator Program.
The pulse pileup simulation code used here for imaging spectroscopy was developed 
by Richard Schwartz, who also helped with the correction for particle contamination.
We thank Kim Tolbert and other \hsi team members for their indispensable technical support.
We also thank Siming Liu for help with modeling particle transport,
Chang Liu and Dale Gary for providing the OVSA microwave spectra,	
James McTiernan for providing his force-free extrapolation of AR 10486,
Yang Liu for discussions of magnetic field measurements. We thank
K. D. Leka, Haisheng Ji, and Tongjiang Wang for various help and discussions.
We are grateful to the referee for critical comments and constructive suggestions.
\soho is a joint project of ESA and NASA.
OVSA is supported by NSF grant AST-0307670 to New Jersey Institute of Technology.

\appendix

\section{A. Effects of Pulse Pileup on Imaging Spectroscopy}
\label{chp1029_pileup}

Pulse pileup refers to the phenomenon that two or more photons close in time are detected as one photon with their
energies summed \citep{SmithD2002SoPh..210...33S}.  When count rates are high, as happens in large flares, 
an artifact appears in the measured spectrum at twice or a higher multiple of
the energy of the peak of the count rate spectrum that is 
at $\sim$6~keV in the \hsi attenuator A0 state, $\sim$10~keV in the A1 state, and $\sim$18~keV in the A3 state.
Due to nonlinear complexity, there is currently no 100\% reliable 
pileup correction algorithm available in the \hsi software, especially for imaging spectroscopy.  
Here we have adopted and improved upon the methods used by \citet[][their \S 2.1]{LiuW2006ApJ...649.1124L}
to estimate the pileup importance and minimize its effects on our analysis.

A general indicator of pileup severity is the livetime, the complement of the deadtime
during which the detector is not able to distinguish among different incident photons.
We find that the fractional livetime averaged over detectors~3--9 and over 4 s intervals
has a V-shaped time profile (not shown) with values $\geq$90\% at the beginning ($<$20:40~UT) 
and end ($>$21:00~UT) of the flare,
and a minimum of 24\% at 20:46:10~UT during the impulsive peak. 
This value is very small compared with the livetime minima of 55\% during the 2002 July 23 X4.8 flare 
and 94\% during the 2002 February 20 C7.5 flare, and indicates severe pileup effects.  
On much shorter time scales, the livetime fluctuates in anti-correlation with the count rate
due to the modulation of the \hsi grids during the spacecraft rotation, and a variation between
5\% and 70\% is found for detector~9 at 20:46:10~UT.
Such fine temporal variations make
pileup correction for imaging spectroscopy even more difficult than for spatially integrated spectroscopy.

\citet{SchwartzRA.pileup.sim.2008AGU} has recently developed a forward-modeling tool to 
simulate pileup effects for imaging spectroscopy, based on input model images consisting 
of sources of user-specified spectra and spatial distributions. This tool has been applied here
to generate simulated images at 20:46:00--20:46:04~UT (near the livetime minimum) using
the real flight instrument response. The model image consists of a thermal elliptic-gaussian LT source
and two nonthermal (power-law) circular-gaussian FP sources, 
whose geometric and spectral parameters were selected to match the observation at 20:46:00--20:46:20~UT. 
We have run simulations for four cases as listed in Table~\ref{chp1029_table_pileup}
with different count rates corresponding to different levels of pileup.
Case A returns the nominal result assuming no pileup at all;
Case C represents this flare with the count rate of CR$=1.1\E{5}$ (counts $\ps$ detector$^{-1}$) 
identical to the maximum of the measured value at 20:46:10~UT.
In each case images in 16 energy bins (used in \S\ref{chp1029_obs_imgspec})
from 6 to 150 keV were reconstructed and then {\it normalized} as if their 
corresponding count rates were identical to the measured value. The final normalization
makes different cases directly comparable and emphasizes the effects of pileup.
%
\begin{table}[thbp]	
\footnotesize  
\caption{Simulated effects of pulse pileup at different levels (indicated by the count rates) on imaging spectroscopy.
The highest count rate during this flare is CR$_{\rm obs}=1.1 \E{5}$ counts $\ps$ detector$^{-1}$.
The fluxes and EMs are {\it normalized} by a factor of CR$_{\rm obs}$/CR, where CR is the count 
rate assigned to each case.
}
\tabcolsep 0.04in	
\begin{tabular}{llrrrcrrrrrr}
\tableline \tableline
Case & Count Rate (CR)             & \multicolumn{3}{c} {FP spectral index}    & &  \multicolumn{3}{c} {FP flux at 50 keV}        & & \multicolumn{2}{c} {LT spectral parameters} \\
     & (cts $\ps$ det$^{-1}$) & \multicolumn{3}{c} {in 50--150 keV} & &  \multicolumn{3}{c} {(ph~$\ps \, \pcms \, \pkeV$)}   & & \multicolumn{2}{c}{in 6--20 keV} \\
                             \cline{3-5}                 \cline{7-9}  	\cline{11-12}  \\  
\multicolumn{2}{l}{(Subscript 1: E-FP, 2: W-FP)}  &  $\gamma_1$   &  $\gamma_2$  &  $\gamma_2-\gamma_1$ & &  $I_1$   &  $I_2$  &  $I_2/I_1$ & & EM ($10^{49} \pcmc$) & $T$ (MK)  \\
\tableline
A & $7.7 \E{4}$ (but no pileup)
                  & 3.71 & 3.82 & 0.11  & &  25.5 & 12.0 & 0.47  & &  15.4 & 30.8 \\ 
B & $1.6 \E{4}$   & 3.74 & 3.85 & 0.11  & &  29.2 & 13.7 & 0.47  & &  15.0 & 30.9 \\ 
C & $1.1 \E{5}$ (measured)  
                  & 3.85 & 3.93 & 0.08  & &  52.8 & 24.0 & 0.46  & &  12.1 & 31.0 \\ 
D & $3.9 \E{5}$   & 3.66 & 3.71 & 0.05  & &  134  & 58.9 & 0.44  & &  6.64 & 31.1 \\ 
\tableline  \end{tabular}
\label{chp1029_table_pileup} \end{table}
%
%
%
 \begin{figure}[thbp]        
 \epsscale{0.5}
 \plotone{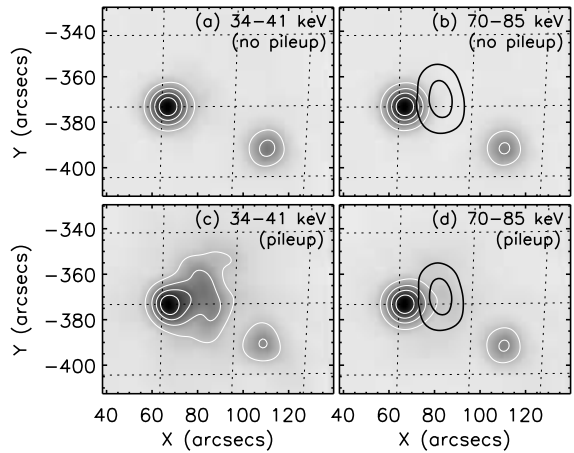}
 \caption[]
 {Simulated images for 20:46:00--20:46:04~UT at different energies.  ({\it a}) and ({\it b}) Case A
 without pileup. ({\it c}) and ({\it d}) Case C with pileup representing this flare.
 The dark contours at 40\% and 80\% of the maximum in ({\it b}) and ({\it d}) represent the 6--8 keV
 image where the LT source dominates.
 }
 \label{chp1029_pileup_img.eps}
 \end{figure}
%
%
 \begin{figure}[thbp]        
 \epsscale{0.39}	
 \plotone{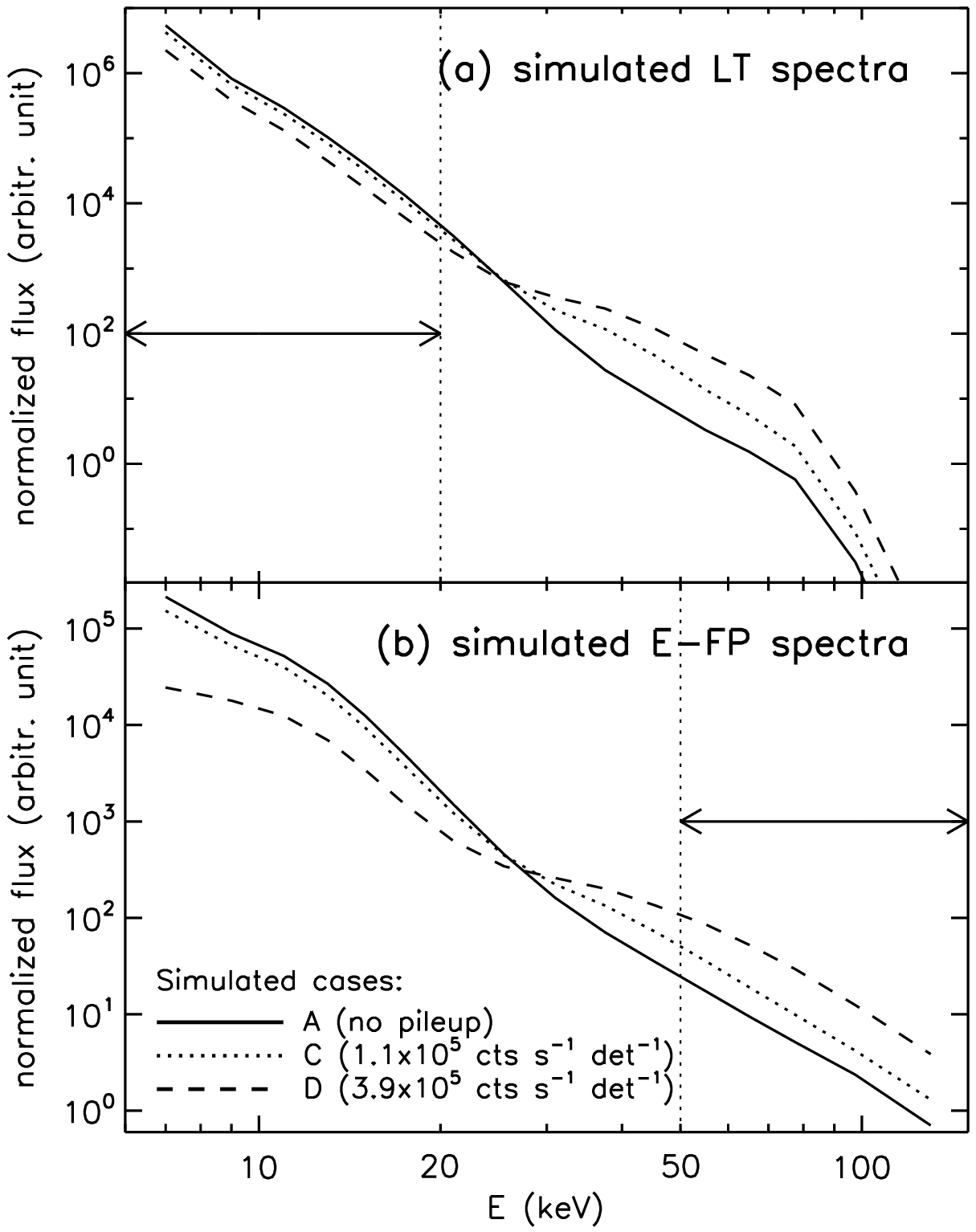}	  
 \plotone{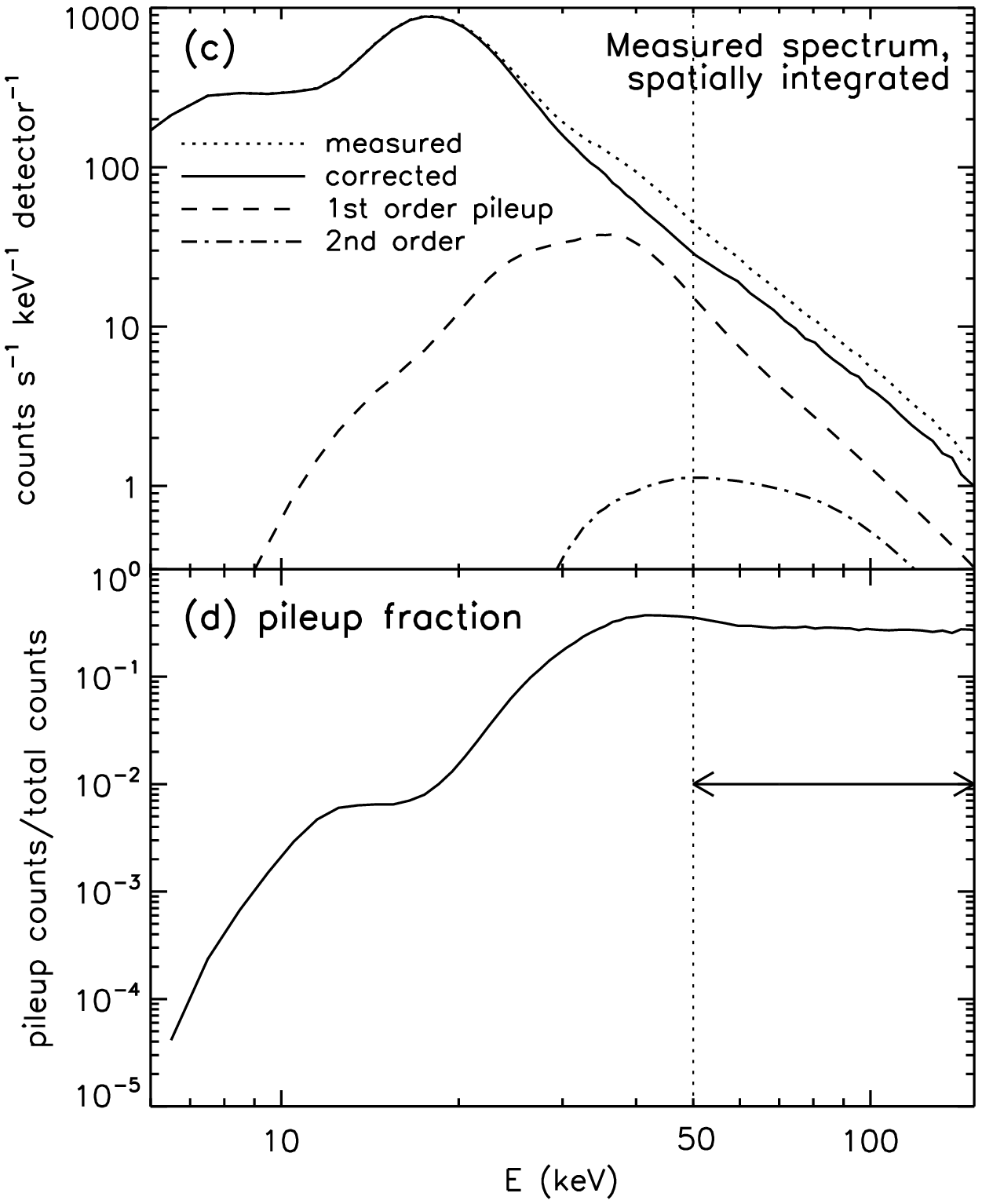}	  
 \caption[]
 { 	{\it Left:} Simulated photon spectra for the LT ({\it a}) 
 and E-FP sources ({\it b}) for Cases A, C, and D at 20:46:00--20:46:04~UT.
 The excessive LT flux $\gtrsim$30~keV and E-FP flux $\lesssim$30~keV are due to the spatial overlap
 of the two sources.	%
   {\it Right:} Measured, spatially integrated, count rate spectrum ({\it c}) averaged over detectors 3--9
 for interval 20:46:00--20:46:20~UT before and after pileup correction \citep{SmithD2002SoPh..210...33S}.
 The broken lines are the estimated first- (two photons) and second-order (three photons) piled-up 
 spectra, with their peaks marked by the vertical dotted lines.
   ({\it d}) The fraction of the total recorded counts that are piled up 
 (including first- and second-order).
 The double arrows indicate the energy ranges used in the analysis.
 }
 \label{chp1029_pileup.eps}
 \end{figure}
%

Let us first examine the effects of pileup on source {\it morphology}. From Figure~\ref{chp1029_pileup_img.eps}
we find the following features in agreement with the speculation of \citet[][their \S 2.1]{LiuW2006ApJ...649.1124L}:
 (1) The source positions, shapes, and sizes are generally well preserved, especially
at low and high energies. This is because photons from the same region within the FWHM resolution of 
a grid have higher probability to pileup as one count in the underlying detector than photons from 
different regions. Therefore during image reconstruction piled-up
photons are likely to be registered back to their original source.
 (2) The main visible artifact is the ``ghost" LT source at intermediate energies (e.g., 34--41~keV).
This is because the count-rate spectrum of this flare (in attenuator state A3) is $\sim$18~keV
at which the LT source dominates over the FP sources, and twice this energy is 36~keV where the first-order
pileup is most pronounce.

Next we check the effects of pileup on source {\it spectra}. In general, pileup shifts photons from low energies
to high energies, as can be seen from Figures~\ref{chp1029_pileup.eps}{\it a} and \ref{chp1029_pileup.eps}{\it b}.
For the LT source, the spectrum is only slightly reduced at low energies ($\lesssim$20~keV), 
with negligible changes in shape, while it is substantially altered in both normalization and shape at high energies. 
The opposite is true for the E-FP source whose spectrum $\gtrsim$50~keV is only elevated
by a roughly constant factor independent of energy (with minimal changes in slope). 
For comparison, we also performed pileup correction to the spatially integrated spectrum
at 20:46:00--20:46:20~UT using the preliminary algorithm of \citet{SmithD2002SoPh..210...33S}
with a tweak factor of 0.6. 
The result shown in Figures~\ref{chp1029_pileup.eps}{\it c} and \ref{chp1029_pileup.eps}{\it d}.
Because the LT source dominates at low energies, while the FPs dominate at high energies, the above
mentioned features at these extreme energies are evident.
The most obvious alteration of the spectrum shape occurs in the 30--40~keV range that covers
twice the peak energy (18~keV) of the count-rate spectrum.

We repeated our imaging spectroscopy analysis here on the simulated data and the results
are summarized in Table~\ref{chp1029_table_pileup}. The striking feature is that the 
{\it spectral shapes} are well preserved across a wide range of pileup severity, as indicated by 
(1) the temperature $T$ inferred from the LT spectra $<$20~keV and
(2) the spectral indexes $\gamma_1$ and $\gamma_2$ and their difference, and the flux ratio $I_2/I_1$
inferred from the FP spectra $>$50~keV. 
The FP portion of these spectral parameters are of our main interest in this paper. 
As to the {\it absolute normalizations}, according to Case C, the estimated FP fluxes can be 
elevated by a factor of 2 and the EM can be reduced by 20\% for this flare.
These are the upper limits of changes as we used here the 
maximum observed count rate at 20:46:10~UT,  	
and the changes are reduced considerably away from this time.
The effects on the FP spectra are further reduced later in the flare as the spatial contamination 
from the LT is alleviated when the FP sources move away.
We conclude that our choice of fitting the LT spectra only below 20~keV and the FP 
spectra only above 50~keV is justified and has effectively minimized pileup effects on our analysis. 
We note in passing that the count spectrum of radiation belt particles estimated in \S~\ref{chp1029_gen_hsi}
appears as a hump peaking at $\sim$50~keV and the counts integrated in the 3--600~keV band 
amount to only $<$8\% of the total counts during 20:40--21:05~UT. Therefore these particles
have a negligible contribution to pileup.


\section{B. Coalignment of Images Obtained by Different Instruments}
\label{chp1029_coalign}

We describe here how we	
coaligned the \hsi images and \soho MDI magnetograms presented in this paper.
(See a general tutorial: 	
  http://hesperia.gsfc.nasa.gov/$\sim$ptg/trace-align.)
{\it RHESSI}'s images are located on the Sun to sub-arcsecond accuracy thanks to its solar limb sensing
aspect system and star based roll angle measurements \citep{FivianM2002SoPh..210...87F}.
MDI images had accurate plate scales and their roll angles were corrected for the solar P-angle, 
but had different absolute origins for the solar $x$ and $y$ coordinates.  
MDI images were corrected to match the \hsi features using the following procedures.



(1)	
 The first step is to identify specific features on the MDI map that have \hsi counterparts.
\citet{Qiu.Gary.BfieldAnomaly2003ApJ.599.615} found good spatial agreement between HXR FPs and
MDI magnetic anomaly features with an apparent sign reversal in a white-light flare.  This was interpreted as
HXR-producing nonthermal electrons being responsible for heating the lower atmosphere, which consequently 
altered the Ni I 6768 \AA\ line profile that is used by MDI to measure the magnetic field.

\begin{figure}[thbp]        
 \epsscale{0.42}	
 \plotone{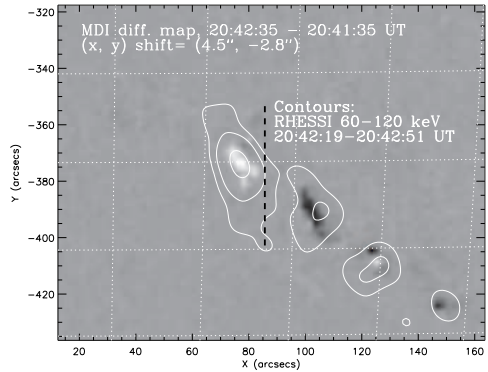}	
 \caption[MDI Running Difference]
 {
 Difference ({\it gray background}) between two MDI magnetograms at 20:42:35 and 20:41:35~UT,
 overlaid with a \hsi 60--120~keV image at 20:42:19--20:42:51~UT (white contours 
 at 15\%, 30\%, \& 80\% of the maximum brightness).
 The MDI map has been corrected to an Earth-centered view and shifted in $x$ and $y$ to 
 best match the four \hsi HXR sources (see text).  The dark dashed line is the same 
 simplified magnetic NL shown in Fig.~\ref{chp1029_HSIonMDI.ps}{\it a}.
 }
 \label{chp1029_mdi_rdiff.eps}
 \end{figure}

We selected two neighboring magnetograms at 20:41:35 and 20:42:35~UT when the magnetic anomaly
features were most pronounced, and subtracted the former from the latter.  This running-differenced 
map (which we call $map_0$, Fig.~\ref{chp1029_mdi_rdiff.eps}) highlights regions of the newest changes, 
which are presumably caused by precipitation of nonthermal electrons and are expected to
appear cospatial with HXR sources.
As evident, there is one (three) apparent increase (decrease) feature(s) mainly in the negative (positive) polarity
(cf., Fig.~\ref{chp1029_HSIonMDI.ps}), which appear as {\it white} ({\it dark}) patches in Figure~\ref{chp1029_mdi_rdiff.eps}. 
Meanwhile, we reconstructed a \hsi image (called $map_1$) at 60--120~keV integrated in the interval
of 20:42:19--20:42:51~UT (a multiple of the \hsi spin period, $\sim$4~s, and closest to 
the corresponding integration time of the second MDI magnetogram, 20:42:20--20:42:50~UT).  We found an one-to-one
correspondence between the four major HXR FP sources (Fig.~\ref{chp1029_mdi_rdiff.eps}, {\it contours}) 
and the magnetic anomalies.

(2)	
We then converted {\it SOHO}'s L1 view to the appearance as viewed from 
an Earth orbit, and we call the resulting differenced MDI magnetogram $map_{0, \,{\rm Earth}}$.

(3)	
Finally, we took the absolute value of MDI $map_{0, \,{\rm Earth}}$ to make a new map called 
$map'_{0, \,{\rm Earth}}$ and coregister it with \hsi $map_1$ by cross-correlation.
The required shifts for $map'_{0, \,{\rm Earth}}$ are $\Delta x=4.5 \arcsec \pm 2.0 \arcsec$ and 
$\Delta y=-2.8 \arcsec \pm 2.0 \arcsec$,
where the $2.0 \arcsec$ uncertainty was estimated with the MDI pixel size.
%
%


\section{C. Derivation of Footpoint HXR Fluxes Resulting from Asymmetric Coronal Column Densities}
\label{chp1029_coldep_math}

Here we derive the numerical expressions for the HXR fluxes of the two FPs and their ratio as a function
of energy resulting from column density asymmetry addressed in \S~\ref{chp1029_asym_coldepth}.
We adopted the empirical expression of \citet[][their eq.~11]{LeachJ1983ApJ...269..715L} for nonthermal 
bremsstrahlung X-ray emission as a function of column density. This expression well approximates 
the Fokker-Planck calculation of particle transport under the influence of Coulomb collisions
that includes energy losses and pitch-angle scattering, the latter of which was neglected in other forms 
of X-ray profiles based on approximate analytical solutions \citep[e.g.,][]{EmslieMachado1987SoPh..107..263E}.
For an injected power-law (index $\delta$) electron flux, the resulting fractional bremsstrahlung 	
emission intensity per unit dimensionless column density $\tau$ at photon energy $k$ 
(in units of rest electron energy $m_e c^2=511$~keV, $m_e$ being the electron mass) can be written as
 \beq
    I_0(\tau, k) = \left(\frac{\delta}{2}-1\right) \left(\frac{k+1}{g k^2}\right) \left(1 + 
                 \tau \frac{k+1}{g k^2}\right)^{-\delta/2}\,,
 \label{ItauEq1} \eeq
where $\tau=N [4 \pi r^2_0 \ln \Lambda] = N/[5 \times 10^{22} \pcms]$ is the dimensionless column density, 
for the classical electron radius $r_0= e^2/m_e c^2=2.8 \times 10^{-13} \cm$ and the Coulomb logarithm
$\ln \Lambda = 20$; $g$ is a factor determined by the pitch-angle distribution of the injected
electron spectrum, which we assumed to be isotropic and thus $g=0.37$ \citep{LeachJ1984PhDT........33L}. 
This emission profile is normalized to unity, 
  $\int_0^\infty I_0(\tau, k)d\tau=1$.
Integrating equation (\ref{ItauEq1}) yields the cumulative photon emission from the injection site ($\tau=0$)
to the transition region ($\tau=\tau_{\rm tr} \equiv N_{\rm tr}/[5 \times 10^{22}$ cm$^{-2}]$,  where 
$N_{\rm tr}= \int_0^{s_{\rm tr}} n[s] ds$ and $s_{\rm tr}$ are the coronal column density and distance to the transition region),
 \beq	
 F_{\unit{Corona}}(\tau_{\rm tr})= \int_0^{\tau_{\rm tr}} I_0(\tau, k) d\tau 
          = 1 - \left(1 + \tau_{\rm tr} \frac{k+1}{g k^2}\right)^{1- \delta/2} \,,
 \label{F_corona_tau_Eq} \eeq
whose complement gives the emission accumulated below the transition region, i.e., the HXR flux
of the FP,
 \beq	
 F_{\unit{FP}}(\tau_{\rm tr}) = \int_{\tau_{\rm tr}}^\infty I_0(\tau, k) d\tau= 1- F_{\unit{Corona}}(\tau_{\rm tr}) 
           = \left(1 + \tau_{\rm tr} \frac{k+1}{g k^2}\right)^{1- \delta/2} \,.
 \label{F_fp_tau_Eq} \eeq
Note that at large photon energies (tens to hundreds of keV), $F_{\unit{Corona}}(\tau_{\rm tr})$ is usually much smaller than
$F_{\unit{FP}}(\tau_{\rm tr})$. In addition, $F_{\unit{Corona}}(\tau_{\rm tr})$ is distributed in a large volume in the leg
of the loop in the relatively {\it tenuous} plasma, while $F_{\unit{FP}}(\tau_{\rm tr})$ is concentrated at the FP
in the {\it dense} transition region and chromosphere.  This results in an even smaller surface brightness 
in the leg than at the FP, which may well exceed the dynamic range of HXR telescopes (e.g., $\gtrsim$10:1
for {\it RHESSI}). This is why leg emission is so rarely observed \citep{LiuW2006ApJ...649.1124L, 
SuiL2006ApJ...645L.157S}.

As we know, a power-law electron flux (index=$\delta$) produces a thick-target (integrated from $\tau=0$ to $\tau=\infty$)
photon spectrum of approximately a power-law, $I_{\unit {thick}}=A_0 k^{-\gamma}$, where $\gamma= \delta -1$ 
\citep{BrownJ1971SoPh, PetrosianV1973ApJ...186..291P} for an isotropically injected electron spectrum, 	
and $A_0$ is the normalization factor [in units of photons~$\ps \, \pcms \, (511 \keV)^{-1}$].
 Since $I_0$ gives the fractional spatial photon distribution at a given energy,
the physical photon spectrum at energy $k$ and at a depth where the overlying column density is $\tau$ can be written as
   $I(\tau, k) = I_{\unit {thick}} I_0(\tau, k)= A_0 k^{-\gamma} I_0(\tau, k)$.
It follows that the X-ray flux of the FP is
  \beq
    I_\unit{FP}(\tau_{\rm tr}, k) =\int_{\tau_{\rm tr}}^\infty I(\tau, k) d\tau
    = I_{\unit {thick}} F_{\unit{FP}}(\tau_{\rm tr})
    = A_0 k^{-\gamma} \left(1 + \tau_{\rm tr} \frac{k+1}{g k^2}\right)^{1- \delta/2} \,,
  \label{Ik_tau_Eq} \eeq
and the photon flux ratio of the two FPs (1 and 2), 
 \beq  
 R_I = {I_{\unit{FP}}(\tau_{\rm tr, 2}, k) \over I_{\unit{FP}}(\tau_{\rm tr,1}, k) }
   = \left(1 + \tau_{\rm tr,2} \frac{k+1}{g k^2}\right)^{1- \delta/2} 
     \left(1 + \tau_{\rm tr,1} \frac{k+1}{g k^2}\right)^{-(1- \delta/2)}\,. 
 \label{RI_tau_Eq} \eeq
The above two equations were used in \S~\ref{chp1029_asym_coldepth} to calculate the FP fluxes 
and their ratio resulting from different coronal column densities.

\section{D. Estimation of Column Densities in Loop Legs}
\label{chp1029_dens_factor}

We describe below the approach to estimate the coronal column densities $N_{\rm tr}$ in the legs of the loop, which is
defined and used in \S~\ref{chp1029_asym_coldepth} as the density integrated along the loop 	
from the acceleration region to the transition region at the FPs. 
In the stochastic acceleration model of \citet{PetrosianV2004ApJ...610..550P}, the LT source is the 
region where particle acceleration takes place \citep{LiuW_2LT.2008ApJ...676..704L, XuY.fwdfit2008ApJ...673..576X}.
Assuming that this picture is true,%
  \footnote{In the competing model of \citet{FletcherL.HudsonHS.Wave2008ApJ...675.1645F}, acceleration by 
  plasma waves takes place near/at the chromosphere. In such a case, our density estimation here must be modified.}	
we thus subtracted%
  \footnote{ 
  There are observations \citep[e.g.,][]{Masuda1994Nature, Aschwanden.t-of-flight.1995ApJ...447..923A}
  suggestive that the acceleration region is located above the SXR loop, 	
  and in that case a distance needs to be added to $l_i$. 
  Such a practice was not attempted here,  
  and as we can see from \S~\ref{chp1029_asym_coldepth}, will not change our conclusions.}
the estimated LT size (i.e., the radius $r$ of the {\it equivalent sphere};
see Fig.~\ref{chp1029_img_spec_time.eps}{\it c}) from the distances along the loop from the LT centroid to 
the FP centroids obtained in \S~\ref{chp1029_corr_space} 
(i.e., $l_i$, where $i$=1 for E-FP and 2 for W-FP; see Fig.~\ref{chp1029_fp_motion.eps}{\it a}), to 
obtain the path lengths in the legs $s_{{\rm tr}, \, i}=l_i-r$. 
Here the FP centroids are assumed to be situated at negligibly small distances below the transition region.
To give the desired column densities $N_{{\rm tr}, \, i}$,		
the path lengths $s_{{\rm tr}, \, i}$ ($i$=1, 2) were then multiplied by the density $n_{\rm leg}$ 
(assumed to be uniform) in the legs of the loop, which was estimated as follows.

The density of the LT source $n_{\rm LT}$ inferred in \S~\ref{chp1029_obs_imgspec} 
(see Fig.~\ref{chp1029_img_spec_time.eps}{\it d}) provides our first guess for the leg density $n_{\rm leg}$
as assumed by \citet{FalewiczR2007A&A}. The relative brightness of {\it nonthermal} bremsstrahlung emission 
the leg and FP provides another important clue.
This is because, for the same reason of collisional losses mentioned in \S~\ref{chp1029_asym_coldepth}, 
the ratio of the leg to FP brightness, particularly at low energies,
is an increasing function of the leg density.  This predicted ratio		
cannot exceed the observed LT-to-FP brightness ratio, because the LT source is where
the maximum loop brightness is located, and it includes additional contributions from  
{\it thermal} emission, piled-up photons, and/or electrons trapped in the acceleration region   
\citep{PetrosianV2004ApJ...610..550P}. This imposes an upper limit for the leg density $n_{\rm leg}$,
which we found to be $n_{\rm leg, \, max}= 0.5 n_{\rm LT}$ based on an error and trial method.
This result indicates that	
the average density in the legs is smaller than the estimated LT density,
or that the LT density is an overestimate due to an underestimate of the volume which could result from 
the lack of knowledge of the source size in the third dimension along the line of sight.
This leg density was then used for column densities $N_{{\rm tr}, \, i}= n_{\rm leg, \, max} (l_i-r)$ 
shown in Figure~\ref{chp1029_vs_time.eps}{\it e}. 



{\footnotesize

}


\end{document}